\documentclass{aastex631}
\pdfoutput=1 
\usepackage{amsmath}
\usepackage[T1]{fontenc}
\usepackage{url}
\usepackage{graphicx}
\usepackage[caption=false]{subfig}
\DeclareMathOperator{\E}{\mathbb{E}}

\usepackage[caption=false]{subfig}

\begin{document}

\title{Rapid localization of gravitational wave sources from compact binary coalescences using deep learning}

\author[0000-0001-8700-3455]{Chayan Chatterjee}
\affiliation{Department of Physics, OzGrav-UWA, The University of Western Australia,\\ 35 Stirling Hwy, Crawley, Western Australia 6009, Australia}

\author[0000-0001-8143-9696]{Manoj Kovalam}
\affiliation{Department of Physics, OzGrav-UWA, The University of Western Australia,\\ 35 Stirling Hwy, Crawley, Western Australia 6009, Australia}

\author[0000-0001-7987-295X]{Linqing Wen}
\affiliation{Department of Physics, OzGrav-UWA, The University of Western Australia,\\ 35 Stirling Hwy, Crawley, Western Australia 6009, Australia}

\author[0000-0002-1481-1993]{Damon Beveridge}
\affiliation{Department of Physics, OzGrav-UWA, The University of Western Australia,\\ 35 Stirling Hwy, Crawley, Western Australia 6009, Australia}

\author[0000-0002-8788-8174]{Foivos Diakogiannis}
\affiliation{The Commonwealth Scientific and Industrial Research Organisation,\\ 7 Conlon St, Waterford, WA, Australia}

\author[0000-0001-5332-3784]{Kevin Vinsen}
\affiliation{International Centre for Radio Astronomy Research, The University of Western Australia\\ M468, 35 Stirling Hwy, Crawley, WA, Australia}



\begin{abstract}
The mergers of neutron star-neutron star and neutron star-black hole binaries are the most promising gravitational wave events with electromagnetic counterparts. The rapid detection, localization and simultaneous multi-messenger follow-up of these sources is of primary importance in the upcoming science runs of the LIGO-Virgo-KAGRA Collaboration. While prompt electromagnetic counterparts during binary mergers can last less than two seconds, the time scales of existing localization methods that use Bayesian techniques, vary from seconds to days. In this paper, we propose the first deep learning-based approach for rapid and accurate sky localization of all types of binary coalescences, including neutron star-neutron star and neutron star-black hole binaries for the first time. Specifically, we train and test a normalizing flow model on matched-filtering output from gravitational wave searches to obtain sky direction posteriors in around 1 sec using a single P100 GPU, which is several orders of magnitude faster than full Bayesian techniques.
\end{abstract}



\section{Introduction} \label{sec:intro}
In 2015, the first direct detection of a gravitational wave (GW) signal from a merging binary black hole (BBH) (\cite{GW150914}) system was made by the LIGO Scientific Collaboration (\cite{LIGO}) and the Virgo Collaboration (\cite{Virgo}). Since then, the total number of GW detections has grown, at the time of writing, to 90 (\cite{GWTC3}), and a new detector, KAGRA (\cite{KAGRA}), has joined the GW search effort. These 90 detections include two confirmed signals from merging binary neutron stars (BNS) and at least four from mergers of neutron star-black hole (NSBH) binaries (\cite{GWTC3}). The first BNS detection in 2017 also involved a coincident observation of a short-gamma ray burst (GRB) originating from the same source (\cite{GW170817, GW170817_1, GW170817_2}), making it the first multi-messenger observation involving both GW and electromagnetic (EM) signals. \\


The observations of these electromagnetic counterparts, including the early kilonova lightcurve, and several other possible emissions in the optical, ultraviolet, X-ray and radio can enrich our understanding of the physics of compact objects, constrain the neutron star equation of state (\cite{X-ray1}), and reveal the processes governing formation and emission of shocks and ejecta (\cite{shock-heated_ejecta}). In GW data analysis, a full Bayesian framework, involving Markov Chain Monte Carlo (MCMC) and nested sampling, contained within the codes BILBY (\cite{Bilby}) and  LAL\begin{small}INFERENCE\end{small} (\cite{LALInference})
has been traditionally used to estimate optimal source parameters of detected GW events (\cite{GWTC3}). These tools are, however, very computationally intensive and time-consuming, since they probe the full 15-dimensional parameter space that typically characterises GW signals. The rapid, Bayesian, non-MCMC code \begin{small}BAYESTAR\end{small} (\cite{Bayestar}) can generate posterior distributions of right ascension ($\alpha$) declination ($\delta$) and luminosity distance in a few seconds. \\ 

Other fast parameter estimation methods use heterodyned likelihood (\cite{NeilCornishpaper}), iterative fitting methods as used in the \texttt{RIFT} algorithm \cite{RIFT} and an excess power approach  (\cite{KippCannonpaper}). What all of these methods have in common is that they are approximations of Bayes theorem:

\begin{equation}
    p(\theta|d) = \frac{p(d|\theta)p(\theta)}{p(d)}, 
\end{equation}
where $p(d|\theta) $ is the likelihood distribution of the observed strain data, assumed to consist of a GW signal $h(\theta) $ plus stationary Gaussian noise $n $. $p(\theta)$ represents our prior belief about the distribution of the source parameters and $p(d) $ is the evidence.






On the other hand, deep learning techniques, in particular, neural posterior estimation (NPE) (\cite{NPE1, NPE2, NPE3}) models have proven to be efficient at estimating accurate posterior distribution of BBH source parameters (\cite{Green, Gabbard, Chua}), enabling sky direction inference at much faster speeds than conventional sampling algorithms like MCMC. 
Previous applications of deep learning for BBH sky localization involved use of a classification approach in which a neural network was trained to learn hand-crafted features from raw GW strain data and associate those features with a particular `pixel' in the sky in which the source may be located (\cite{Chatterjee2019}). The prediction of the neural network is a probability distribution of the source's sky direction over those pixels. While this approach showed promising results for simulated BBH signals over 2048 sky pixels, it does not scale well for any arbitrary number of pixels, leading to, in some cases, poor localization due to low angular resolution.  \\  

Moreover, there have been no published results, at the time of writing, on deep learning-based posterior estimation of BNS and NSBH sources, which are the primary compact binary coalescence (CBC) candidates with potential EM counterparts. The main challenge involving the use of deep learning models for BNS and NSBH parameter estimation is that the GW waveforms of these sources are much longer than BBHs and can last up to a thousand seconds at detector design sensitivity. This means that for the same overall signal-to-noise ratio (S/N), BNS and NSBH GW signals will have lower signal strength in each individual time segment, compared to BBH. This makes feature extraction using deep learning models for BNS and NSBH strains extremely challenging and is one of the possible reasons why no parameter estimation results on BNS and NSBH using deep learning has been demonstrated yet. \\

In this paper, we introduce \texttt{GW-SkyLocator}, the first deep learning model for sky localization of all compact binaries -- BBH, BNS and NSBH, at orders of magnitude faster speeds than standard Bayesian techniques. We have benchmarked the performance of our model against \begin{small}BAYESTAR\end{small} and other standard samplers like  BILBY and  LAL\begin{small}INFERENCE\end{small} using both real GE events and simulated injections in Gaussian noise. This paper is organized as follows. In Section 2, we discuss the novelty of our deep learning-based sky localization method, in Section 3, we provide the details of our data generation process and describe our neural network architecture. In Section 4, we compare the performance of \texttt{GW-SkyLocator} against \begin{small}BAYESTAR\end{small}, BILBY and  LAL\begin{small}INFERENCE\end{small} for simulated injections and real events. We summarize our findings in Section 5 and discuss plans for future work. \\

\section{Method}

\begin{figure}
\begin{center}
     \includegraphics[scale=0.35]{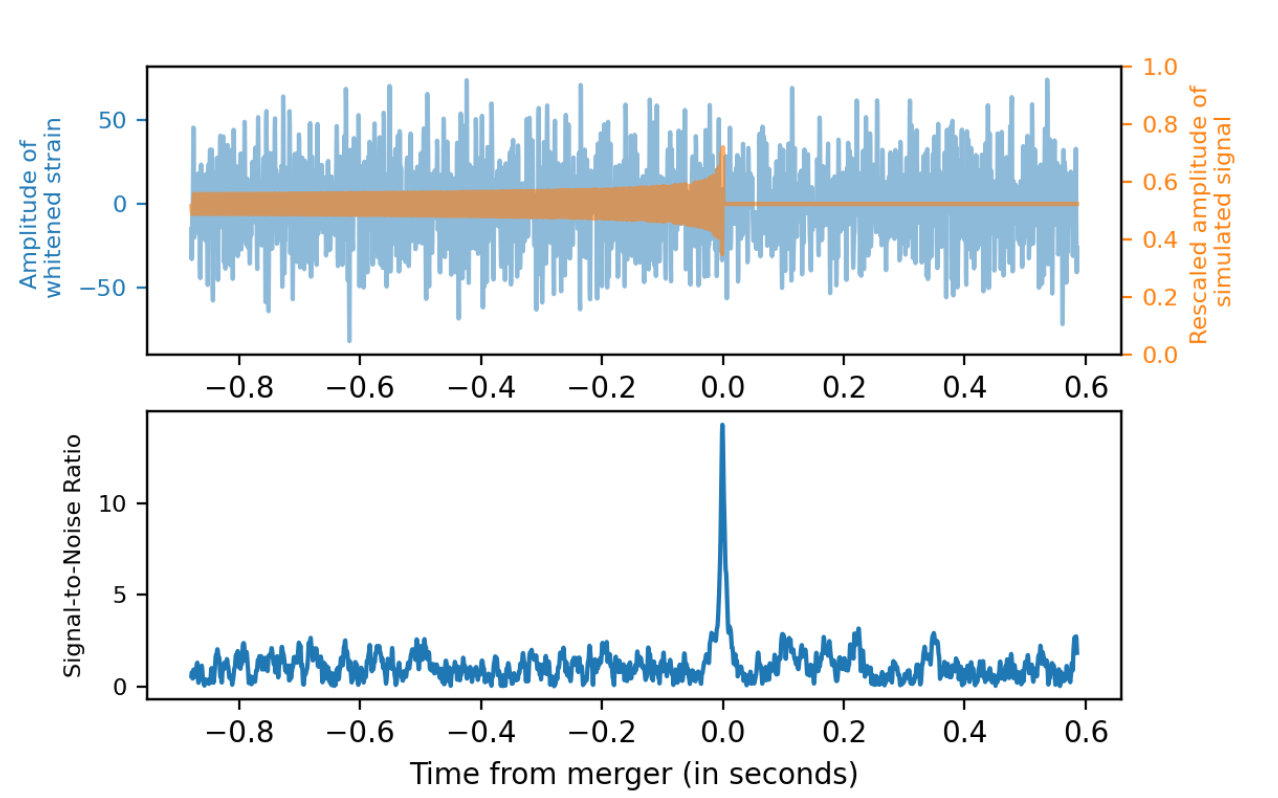}
\caption{A BNS waveform, the strain signal obtained after injecting the waveform in noise, and the corresponding S/N time series obtained by matched filtering the strain with the optimal waveform template is shown. Top panel: The whitened GW strain signal corresponding to a BNS merger is plotted in blue. This data was obtained after injecting the GW waveform (rescaled amplitude shown in orange) into stationary Gaussian noise coloured by advanced LIGO PSD. Bottom panel: The optimal S/N time series obtained by matched filtering the GW strain with the signal template shown in the above panel.}
\label{fig:1}
\end{center}
\end{figure}

Previous deep learning-based methods for parameter estimation used the GW strain data directly for training and testing. Here, we have adopted an alternative approach, first introduced by \begin{small}BAYESTAR\end{small} (\cite{Bayestar}), in which the S/N time series data is used instead of the full GW strains for rapid sky localization. We train our model, called \texttt{GW-SkyLocator} on the S/N time series data, obtained by the template-based matched filtering technique used for detecting modeled GW signals (\cite{Maggiorie_book})
in existing online and offline GW search pipelines  (\cite{SPIIR, GSTLAL, MBTA, PyCBCpipeline, cWB}). Our model also predicts the posterior distribution of the sky location over any arbitrary number of sky pixels, and supports generation of multi-order skymaps, thus alleviating the problems associated with the approach in Chatterjee et al. (2019) (\cite{Chatterjee2019}). The S/N time series of a GW signal can be thought of as a `condensed' version of the GW strain data, in which all information about the GW source sky direction is preserved in its amplitude and phase (\cite{ShaunHooper}), with a  typical length of 0.1 sec around the merger time. The top panel in Fig.~\ref{fig:1} shows an example BNS waveform and the corresponding strain obtained after injecting the signal in stationary Gaussian noise with advanced LIGO Power Spectral Density (PSD). The bottom panel shows the S/N time series obtained by matched filtering the strain data with the signal template. The time difference of the S/N peaks between pairs of detectors represents the arrival time delays of the signals, which is essential information for source localisation based on the triangulation method (\cite{Wen&Chen}). Since the sensitivity of each detector to incoming GW signals as a function of the sky direction of the source, the relative amplitudes and phases of the S/N time series at peak, which depend on the interferometers' directional response to GWs, further helps constrain the possible sky directions (\cite{ShaunHooper}). The S/N time series is also readily available from matched filtering-based GW search pipelines, which means we can directly apply our model in conjuction with an online GW detection pipeline to perform rapid sky localization immediately after detection. \\

Matched filtering involves cross-correlation of the GW strain data, $d$, with waveform templates $h(\theta)$ that depend on the intrinsic source parameters $\theta$, i.e., masses and spins. The output of matched filtering is the S/N time series $\rho (t)$, defined as (\cite{ShaunHooper}), 



\begin{equation}
    \rho(t) = \frac{z(t)}{\sigma},
\end{equation}

where $z(t) $ is the complex matched-filter defined as (\cite{ShaunHooper}),

\begin{equation}
    z(t) = 2 \left(  \int_{-\infty}^{\infty}\frac{\widetilde{d}(f)\widetilde{h}^*_{c}(f)}{S_{n}(|f|)} + i\int_{-\infty}^{\infty}\frac{\widetilde{d}(f)\widetilde{h}^*_{s}(f)}{S_{n}(|f|)} \right) e^{2 \pi ift} df,
\end{equation}

Here $\widetilde{d}(f)$ is the Fourier transform of the GW strain, $\widetilde{h}_{c}(f)$ and $\widetilde{h}_{s}(f)$ are the Fourier transforms of the cosine and sine components of the templatse used for matched filtering, and $S_{n}(f)$ is the PSD of the noise. $\sigma $ is the standard deviation of the real part of the matched filter, which is used to normalize $\rho(t) $. \\
  
In this work, we train our model on the `optimal' S/N time series data, $\rho^{i}_{\text{opt}}(t)$, and the intrinsic parameters $\hat{\theta}_{in}^{i, \text{opt}}$ of the matched filter used to generate $\rho^{i}_{\text{opt}}(t)$. Here the index $i $ refers to the $i$th training sample. This matched filter, having the same source parameters as the injection, optimally extracts the signal from the noise with the highest S/N. Matched filtering with non-optimal templates, i.e., templates having different parameters compared to the injection, produces lower S/N output. During testing, we make direct comparisons of our model with \begin{small}BAYESTAR\end{small}. We first generate 2000 injections each for BBH, BNS and NSBH sources using parameters described in the Methods section, and then run a simulated matched filtering pipeline to generate triggers. Samples having S/N $>$ 4 in atleast two detectors and a total network S/N between 9 and 40 are then picked for inference using \texttt{GW-SkyLocator} and \begin{small}BAYESTAR\end{small}, and the results compared. In the future, we plan to run an end-to-end GW search pipeline on both Gaussian and real noise injections and test our model's performance on the recovered events. These investigations will directly inform our method's scalability and robustness for online sky localization in future LVK observation runs. \\

The exact relation between $\rho_{\text{opt}}$ and sky co-ordinates $\alpha $ and $\delta $ is given in eq. 7.181 in (\cite{Maggiorie_book}). Since our model is trained and tested on very short (0.2 sec long) $\rho_{\text{opt}}$(t) data, we are able to obtain highly accurate localization results for all CBC sources with a relatively simple neural network architecture, compared to deeper and more sophisticated neural networks used in similar analyses (\cite{Green, Gabbard}). The use of short S/N time series data instead of strains for training and inference was the crucial reason for the success of our model on BNS and NSBH events, since machine learning models usually struggle to extract features from hundreds of seconds long input data, even after robust feature engineering, data compression or the use of extremely deep neural networks. \\

Our model consists of three main components: (i) a ResNet-34 network (\cite{ResNet-34}) to extract essential features from the complex S/N time series data $\rho^{i}(t)$, (ii) a Normalizing Flow \cite{NormalizingFlow} model, in particular, a Masked Autoregressive Flow (MAF) network (\cite{MAF1, MAF2}), which generates the posterior distribution of $\alpha $ and $\delta $ given the input data, $p_{true}(\alpha, \delta|\rho^{i}(t), \hat{\theta}_{in}^{i})$, and (iii) a multi-layered fully connected neural network that extracts features from the intrinsic parameters, $\hat{\theta}_{in}^{i} $. The features extracted from the ResNet-34 and fully-connected networks are combined and passed as a conditional input to the MAF. We condition the Normalizing Flow model on the intrinsic parameters of the templates, in addition to the S/N time series, in order to provide additional context to our network during training and inference. This is particularly motivated by \begin{small}BAYESTAR\end{small}'s algorithm, which achieves a much higher inference speed-up over parameter estimation codes like LALI\begin{small}NFERENCE\end{small}, by fixing the intrinsic parameters to values near the peak of the likelihood, rather the marginalizing over them, as is done in offline parameter estimation. We found that including the intrinsic parameter information helped the model constrain the sky position posterior better as they encode information about the shape of the S/N time series peak around merger. While testing our model on real GW events, as discussed in later sections, we use the intrinsic parameters of the best-matched templates that produce the highest S/N output, as for such cases the exact intrinsic parameters are unknown. \\

A Normalizing Flow is a chain of invertible transformations that maps a simple base distribution, like a multi-variate Gaussian, to a more complex target distribution. During training, the model learns to map samples from the base distribution into the target posterior samples. If $z $ is a random sample drawn from the base distribution $\pi(z) $ and $f $ is the transformation induced by the Normalizing Flow, then the new random variable is $x = f(z)$. Since $f $ is invertible, we can write $z  = f^{-1}(x)$. The probability distribution of the new random variable $p(x) $ is therefore given by the change of variables formula for probability distributions:

\begin{equation}
    p(x) = \pi(f^{-1}(x))\left|\text{det}\frac{df^{-1}}{dx}\right|,
\end{equation}

During testing, the trained network can not only generate samples from the target distribution in milli-seconds, but also directly evaluate the probability density of the data. \\ 

We use a multivariate Gaussian as the base distribution, and the MAF as the transformation $f $, to obtain our target posterior $p_{\phi}(\alpha, \delta| \rho^{i}(t), \hat{\theta}_{in}^{i})$. Since our target distribution is a conditional probability distribution, the transformation $f $ is conditioned on features extracted from $\rho^{i}(t)$ and $\hat{\theta}_{in}^{i}$ using the ResNet-34 and fully connected networks. The architectures of these networks are discussed in the Methods section. \\

The MAF models a joint probability density by decomposing it into a product of one-dimensional conditional densities. This is implemented efficiently using a fully connected neural network with a special architecture called Masked Autoencoder for Distribution Estimation (MADE) (\cite{MADE}). The characteristic of the MADE network is that the connections between several neurons in the network are `masked' or set to zero. This preserves the `autoregressive' property of the network which is necessary for correctly modelling the one-dimensional conditional densities at the output layer. \\

\begin{figure}
\begin{center}
  \includegraphics[scale=0.65]{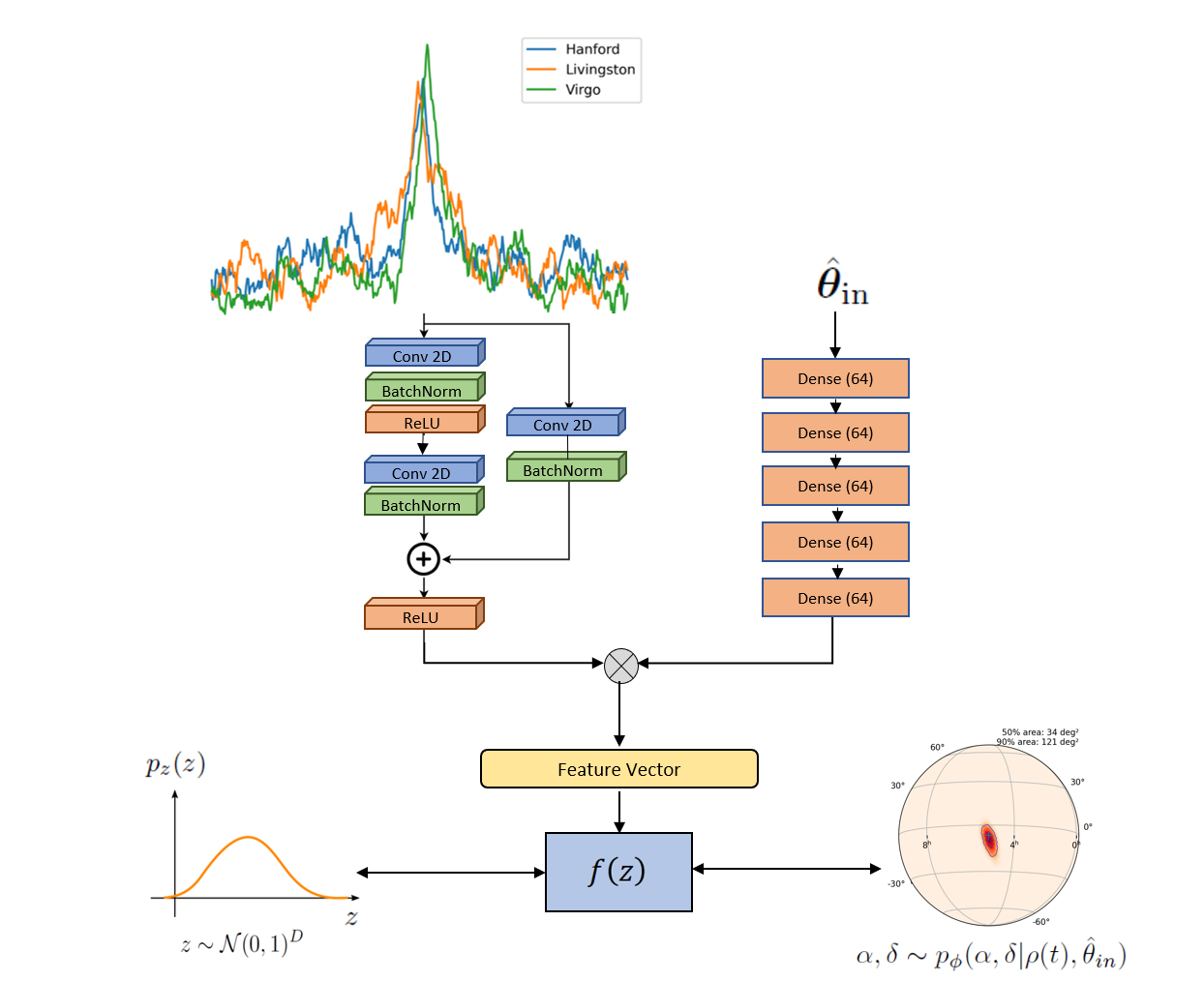}
\caption{\label{fig:Model_architecture} Architecture of our model, \texttt{GW-SkyLocator}. The input data, consisting of the S/N time series, $\rho(t) $ and intrinsic parameters, $\hat{\theta}_{in}$ are provided to the network through the ResNet-34 channel (only one ResNet block is shown here) and the multi-layered fully connected (Dense) network channel respectively. The features extracted by $\rho(t) $ and $\hat{\theta}_{in}$ are then combined and provided as conditional input to the MAF, denoted by $f(z)$.}

%
\end{center}
\label{fig:2}
\end{figure}

Fig.~\ref{fig:2} shows the workflow of the \texttt{GW-SkyLocator}. The MAF takes as input random samples ($z $) drawn from our base distribution, which we choose to be a multivariate Gaussian, and transforms them into samples of our target distribution. The $\alpha $ parameter is cyclic, i.e. for a fixed $\delta $, $\alpha = 0 $ and $\alpha = 2 \pi $ points to the same location in the sky. In order to make our model learn this cyclical nature of $\alpha $, and prevent it from mistaking $\alpha = 0 $ with $\alpha = 2 \pi $ and vice-versa, we decompose $\alpha $ into $\alpha_{\text{x}}$ and $\alpha_{\text{y}} $, equal to the cosine and sine components of $\alpha $ respectively. Therefore, instead of predicting $\alpha $ and $\delta $ directly, we predict $\alpha_{\text{x}}$, $\alpha_{\text{y}} $ and $\delta $. We found that doing this trick improved the performance of the model on injection samples located at $\alpha = 0 $ and $2\pi $. The final $\alpha $ posterior is obtained by taking the arc tangent of $\text{cos}\, \alpha $ and $\text{sin}\, \alpha$. The Normalizing Flow model then scales and shifts the $z $ samples to obtain $p_{\phi}(\alpha, \delta | \rho_{\text{opt}} $), the model's approximation of $p_{\text{true}}(\alpha, \delta | \rho_{\text{opt}}) $. \\





Training a neural network involves minimizing an objective or loss function, which for our case is the expectation over $\rho(t) $  of the cross entropy between the true posterior distribution $p_{true}(\alpha, \delta|\rho(t),\hat{\theta}_{in})$ and the prediction of our model, $p_{\phi}(\alpha, \delta|\rho(t), \hat{\theta}_{in})$. This is given by:

\begin{equation}\label{eq:6}
    L = \int d\rho\, d\hat{\theta}_{in} \, p_{true}(\rho(t),\hat{\theta}_{in}) \int d\theta p_{true}(\theta|\rho(t),\hat{\theta}_{in})
    \text{log}\, p_{\phi}(\theta|\rho(t),\hat{\theta}_{in}),
\end{equation}

where $\theta = (\cos(\alpha), \sin(\alpha), \delta)$. 

Using Bayes' theorem and Monte Carlo approximation, we obtain the following approximate expression for the loss function:

\begin{equation}
    L \approx -\frac{1}{N}\sum_{i=1}^{N}\text{log}\, p_{\phi}(\theta^{(i)}|\rho^{(i)}(t),\hat{\theta}_{in}),
\end{equation}

where $N $ is the size of a `mini-batch' or subset of the full training set. Equation (6) shows that the optimization step is free from costly likelihood and posterior evaluations. Instead, we only need to draw samples from the likelihood, which is readily available as S/N time series outputs from online GW search pipelines (\cite{SPIIR, GSTLAL, MBTA, PyCBCpipeline, cWB}).
The optimization of $L $ is done using stochastic gradient descent, specifically using the Adam optimizer (\cite{adam}). \\

Under the transformation given in eq. 4, the loss function, eq. 6 becomes:

\begin{equation}
     L = \E_{p_{\text{true}}(\theta)}\E_{p_{\text{true}}(\rho(t),\hat{\theta}_{in}|\theta)}\Bigg[-\text{log}\, \mathcal{N}(0,1)^{n}(f^{-1}(\theta))
    - \text{log} \left| \text{det} \frac{\partial(f_{1}^{-1}, ... , f _{n}^{-1})}{\partial(\theta_{1}, ..., \theta_{n})} \right|\Bigg],
\end{equation}

where $n $ refers to the number of parameters/labels being predicted.

\section{Data Generation}

We trained and tested \texttt{GW-SkyLocator} on 5 $\times$ 10$^5$ and 2000 injections respectively for each of the three CBC sources - BBH, BNS and NSBH. For each injection, the GW strain was generated by injecting a simulated waveform in stationary Gaussian noise with advanced LIGO Zero Detuned High Power PSD. Matched filtering was then performed on the strains to obtain the S/N time series. For the training set ($\rho_{\text{opt}}(t)$, $\hat{\theta}^{\text{opt}}_{in}$), matched filtering was done using the optimal filters, as explained earlier. The details of the prior distributions of the injection parameters of our training and test sets are discussed in the Methods section. For each CBC class, the same network architecture was used, which shows the robustness of our approach. During inference, we generate 5000 $\alpha $ and $\delta $ samples from the trained model and evaluate a Gaussian kernel density estimator (KDE) over the samples with a fixed bandwidth of 0.02. The trained KDE is then used to evaluate the probability density over sky pixels using the adaptive sampling scheme employed by \begin{small}BAYESTAR\end{small}, in which the posterior is first evaluated over 16 HEALPix grids (\cite{HEALPix}), which corresponds to a single sky grid area of 13.4 deg$^{2}$. The highest probability grids are then adaptively subdivided into smaller grids over which the posterior evaluation is repeated. This process is repeated seven times to obtain the final multi-order sky map for each test sample. We then run \begin{small}BAYESTAR\end{small} on these injections and compute the areas of different credible intervals for comparison. \\

Training \texttt{GW-SkyLocator} takes $\sim $ 1.5 hours on a P100 GPU. Generating 5000 samples from the trained model during inference takes a few milli-seconds on the same computational resource. The total inference time, including generation of samples, fitting the KDE and evaluating the probability density over the adaptive mesh takes around $\sim $ 1.2 secs on average. \\

The injected signals used in this analysis are generated using the \texttt{LALSuite} package (\cite{LALSuite}). For BBHs, the source-frame primary mass distribution is drawn from a Salpeter initial mass function (IMF) distribution (\cite{Salpeter}), with values lying between 10 and 80 M$_\odot$, and the secondary is drawn from a uniform distribution between 10 M$_\odot$ and the value of the primary mass. For BNS, both components are drawn from a uniform distribution between 1.0 and 2.5 M$_\odot$. For NSBH, the neutron star mass is drawn from a uniform distribution between  1.0 and 2.5 M$_\odot$ and the black hole is drawn from the Salpeter IMF between 10 and 80 M$_\odot$. For the test sets, the BBH injections are simulated with uniform distribution in comoving volume up to a redshift of 2.5, while for BNS and NSBH, the injections are uniform in redshifts up to 0.35 and 0.25 respectively. These distributions are similar to what has been used by the LVK Collaboration for analysis of both simulated and O3 data, but with a narrower black hole mass range between 10 and 80 M$_\odot$ (\cite{GWTC3,Messick,Kovalam,Sachdev}). \\

We performed matched filtering on the simulated strains to obtain the S/N time series, and discarded samples with network S/N $<$ 8 and $ >$ 40. While signals with network S/N $<$ 8 are not considered significant detections, ones with S/N $>$ 40 were discarded because such loud events are expected to be extremely rare, even at the design sensitivities of the ground-based GW interferometers. In a future work, we plan to train different versions of \texttt{GW-SkyLocator} with signals having different S/N ranges like 8 to 20, 20 to 40, 40 to 60 and so on. This strategy is expected to yield better results than what we may obtain from training a single model, since the variance of the S/N distribution in each individual training set distribution is smaller than in a single training set covering the entire S/N range. \\

For generating the training set, instead of simulating signals with similar redshift distribution as the test sets, we adopt a uniform network S/N distribution between 8 and 40 for all three cases. This is done to ensure there are sufficient number of training samples at each S/N bin and therefore avoid potential biases in the network's performance arising from training our model with a non-uniform S/N distribution. For BBH samples, we use the waveform approximant \texttt{SEOBNRv4} (\cite{SEOBNRv4}) and \texttt{SpinTaylorT4} (\cite{SpinTaylorT4}) for BNS and NSBH. All the waveforms were simulated with a fixed coalescence GPS time of 1187008882.4.
The generated waveforms are then injected into Gaussian noise and the resultant strains, sampled at 2048 Hz, are whitened with an estimate of the PSD made using Welch median average method (\cite{Welch_method}) on simulated data. For generating the training samples, after whitening, we rescale the amplitudes of the strains to match a desired network S/N. In order to do that, we first calculate the network optimal matched filtering S/N (NOMF-S/N) of the whitened waveforms using the formula:

\begin{equation}
    \text{NOMF-S/N} = \sqrt{\text{S/N(H1)}^2 + \text{S/N(L1)}^2 + \text{S/N(V1)}^2}
\end{equation}

We then use the NOMF-S/N to calculate a scaling factor that we use to re-scale the strain amplitudes to match the injection S/N drawn from the prior network S/N distribution. After rescaling the strains, matched filtering is performed using the optimal templates to obtain $\rho_{\text{opt}} (t)$. We chop $\rho_{\text{opt}}(t)$ for all CBC sources to 0.2 secs around the merger peak. Sample generation and matched filtering was implemented with a modified version of the code developed by (\cite{Gebherd}) that uses \texttt{PyCBC} software (\cite{PyCBC}). \texttt{GW-SkyLocator} was written in TensorFlow 2.4 (\cite{TensorFlow}) using the Python language.

\subsection{Masked Autoregressive Flow}

In this section, we review Masked Autoregressive Flows and the autoregressive density estimation technique. The MAF is a type of normalizing flow that maps a simple base distribution to a more complex distribution. In autoregressive density estimation techniques, a model learns a complex joint density by decomposing it into a product of one-dimensional conditional densities as follows (\cite{MAF1, MAF2}):

\begin{equation}
    p(x) = \prod_{i}^{n} p(x_{i}|x_{1:i-1}),
\end{equation}

In autoregressive models, these conditional densities are usually parameterized as a univariate Gaussian. The means and standard deviations of the Gaussians are predicted by neural networks that depend on the previous $x_{1:i-1} $:

\begin{equation}
    p(x_{i}|x_{1:i-1}) = \mathcal{N}(x_{i}|\mu_{i}, \text{exp}(\alpha_{i})),
\end{equation}

where,

\begin{equation}
    \mu_{i} = f_{\mu_{i}}(x_{1:i-1}),
\end{equation}
and 
\begin{equation}
    \alpha_{i} = f_{\alpha_{i}}(x_{1:i-1}),
\end{equation}
for i = 1,...,$n $.

$f_{\mu_{i}}$ and $f_{\alpha_{i}} $ are the transformations induced by the MAF. If we sample $u \sim \mathcal{N}(0,1)^{n} $, then,

\begin{equation}
    x_{i} = \mu_{i}(x_{1:i-1}) + \mu_{i}\text{exp}\,\alpha_{i}(x_{1:i-1}), 
\end{equation}
defines a sample from $p(x) $.

This defines a transformation of the base distribution, which is a univariate normal, to the target distribution $p(x) $. We can stack multiple autoregressive transformations to create a sufficiently expressive flow. It is easy to compute the inverse transformation $u = f^{-1}(x) $ since it only requires us to reverse the scale and shift as follows:

\begin{equation}
    u = (x-f_{\mu}(x))/\text{exp}\,(f_{\alpha}(x))
\end{equation}

Therefore the forward and inverse pass of the flow both require only the forward evaluation of $f_{\mu}$ and $f_{\alpha}$. The Jacobian in eq. 7 can be calculated as follows:

\begin{equation}
    \left| \text{det}\, \frac{\partial(f_{1}^{-1}, ..., f_{n}^{-1})}{\partial(x_{1}, ... ,x_{n})} \right| = \text{exp}\,(-\sum_{i=1}^{n}\alpha_{i}(x_{1:i-1})),
\end{equation}

For parameterizing a Normalizing Flow, two conditions need to be necessarily fulfilled: the inverse of the transformation must exist and the Jacobian must be easy to compute. The above arguments suggest that the MAF satisfied both the requirements necessary for a normalizing flow. The autoregressive property is implemented in neural networks by `masking' the weights of a fully connected neural network, or by leaving out only those connections which are allowed, with the rest being set to zero. This architecture is called the MADE or Masked Autoencoder for Density Estimation. Usually to obtain a sufficiently flexible model, a stack of several MADE networks are used. Between each pair of MADE networks, the order of the components are permuted so that if one network is unable to model a distribution well due to poor choice of variable ordering, a subsequent network with a different combination of weights and masking is able to do it. In our MAF implementation, we also use a batch normalization layer between each MADE unit to stabilize training and prevent the performance of the model from suffering due to spurious random weight initializations. 

\subsection{Network Architecture}

In this section, we describe the architectures of the different components of \texttt{GW-SkyLocator}. We use a ResNet-34 network (\cite{ResNet-34}) and a multi-layered fully connected network to extract features from the complex S/N time series data and the intrinsic parameters of the templates used for generating them using matched filtering. The ResNet-34 is a deep network with 34 layers which has a special feature called the skip connections. The skip connection connects the output of one layer to another by skipping some layers in between. These layers, together with the skip connection forms a residual block. ResNets are made by stacking these residual blocks together. The advantage of using several skip connections is that the network starts to learn the features of the input data right from the start of training, even if some layers have not started learning, i.e., their weights are close to the values they were initialized with (usually close to zero). The skip connections make it possible for the signal to flow through the whole network and prevent training degradation and the vanishing gradients problem usually observed in very deep neural networks. \\

The first two layers of the ResNet-34 network are 2D convolutional layers with ReLU activation function and a batch normalization layer. We use 32 filters for the first convolutional layer with a kernel size of 7 and stride of 2. The stride determines the length by which the receptive field of a convolutional kernel shifts over the activations of the previous layer. We then use a max-pooling layer with a pool size of 3 to reduce the input dimensions by a factor of 3. This is followed by a stack of residual units. After every few residual units, the number of filters is doubled, starting from 32. Each residual unit is comprised of two convolutional layers with batch normalization, ReLU activation and kernel size of 3. Overall, our ResNet-34 implementation consists of 34 layers (only counting the convolutional and fully connected layers) containing 3 residual units with 32 filters, followed by 4 residual units with 64 filters, followed by 6 residual units with 128 filters, and then 3 residual units with 256 filters. The intrinsic parameters are passed through a stack of five fully connected layers with ReLU activation function. The features from the final layers of the ResNet and fully connected networks are then combined into a 1d feature vector and passed as conditional input to the MAF. We use a stack of 10 MADE units, each consisting of 5 hidden layers with 256 neurons. In between each pair of MADE networks, we use batch normalization to stabilize training and produce a more flexible model. This architecture is consistent across all the three CBC sources we have considered in this study. We arrived at this architecture from extensive tests and experimentation over different network hyperparameters like number of residual and MADE blocks, number of hidden layers in the MADE blocks, and the number and size of kernels in the convolutional layers. These hyperparameter choices gave us the best results for injection runs conducted in Gaussian noise with minimal overfitting during the training, validation and inference steps.

\section{Results}

\subsection{Injection run on simulated Gaussian noise} 
\begin{figure}
\gridline{\fig{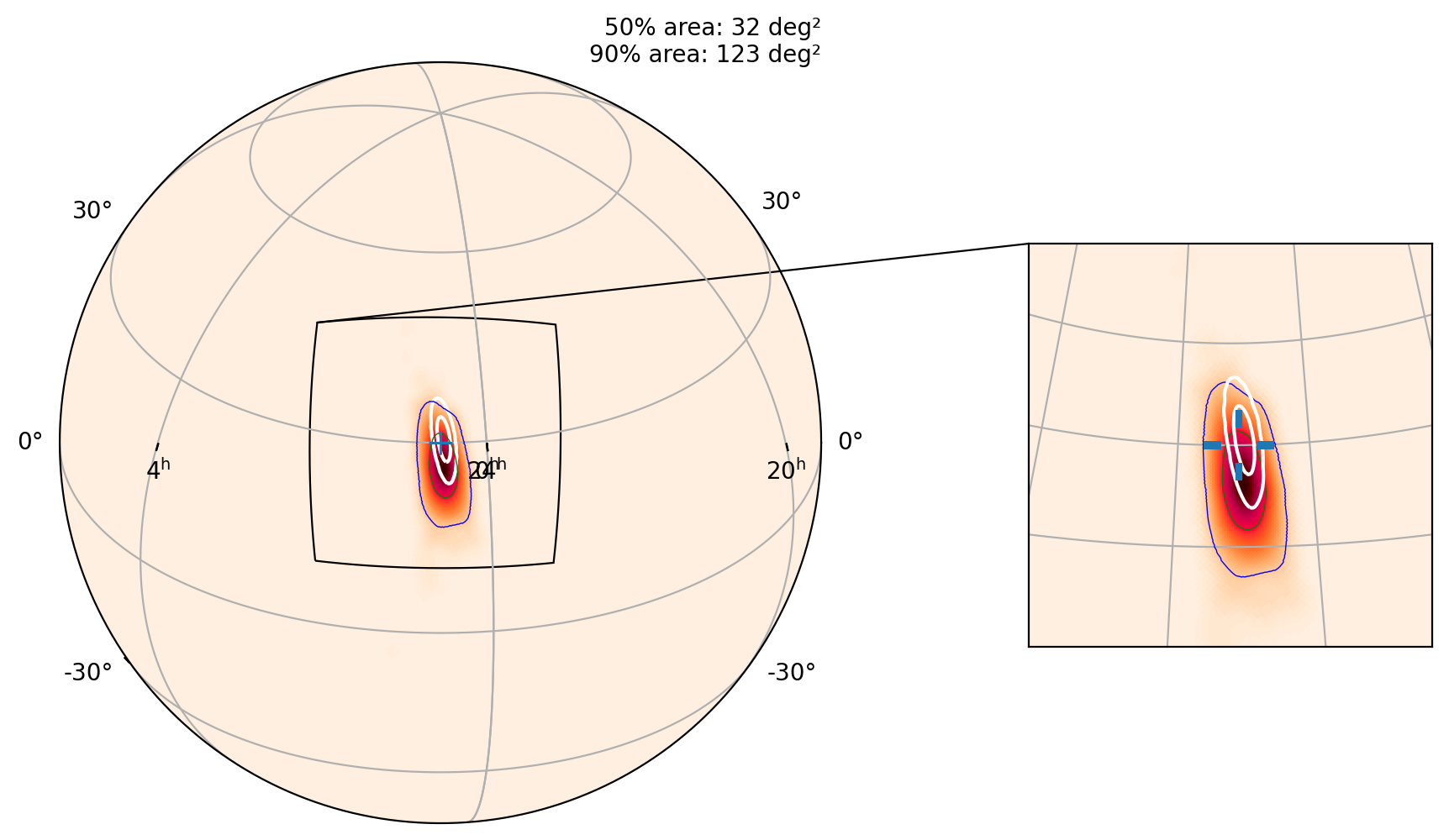}{0.45\textwidth}{(a)}
          \fig{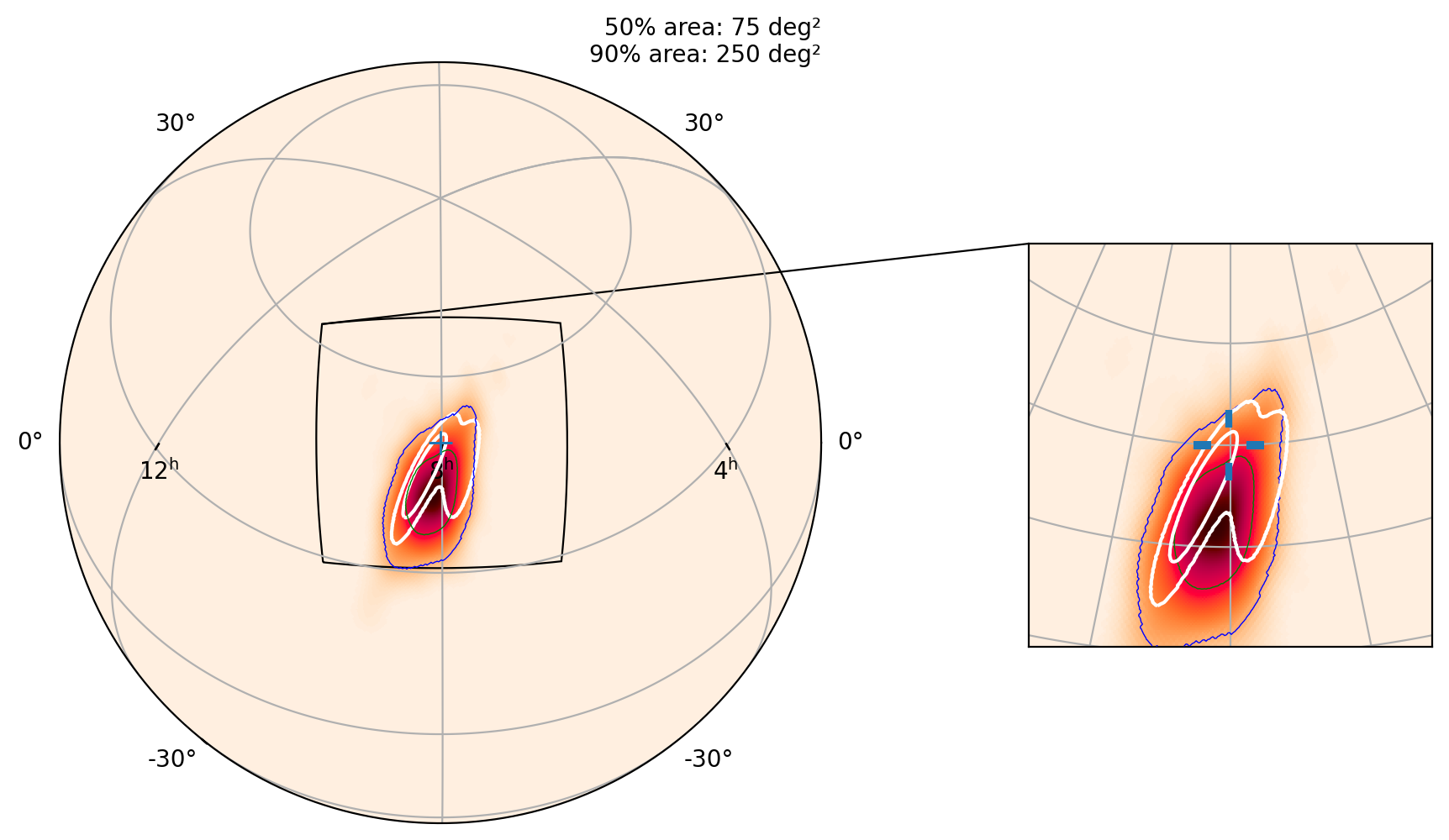}{0.45\textwidth}{(b)}}
\gridline{\fig{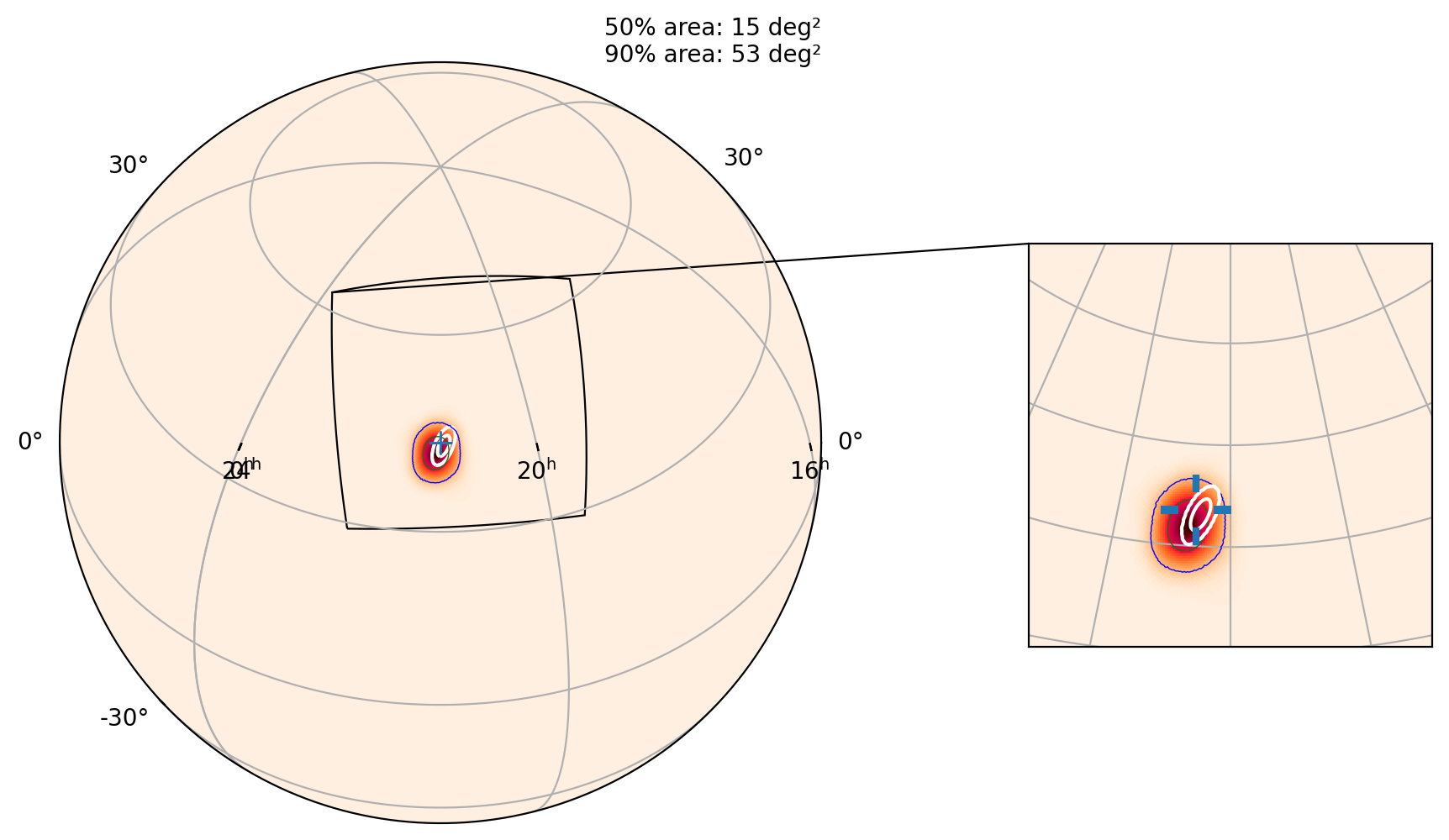}{0.45\textwidth}{(c)}}
\gridline{\fig{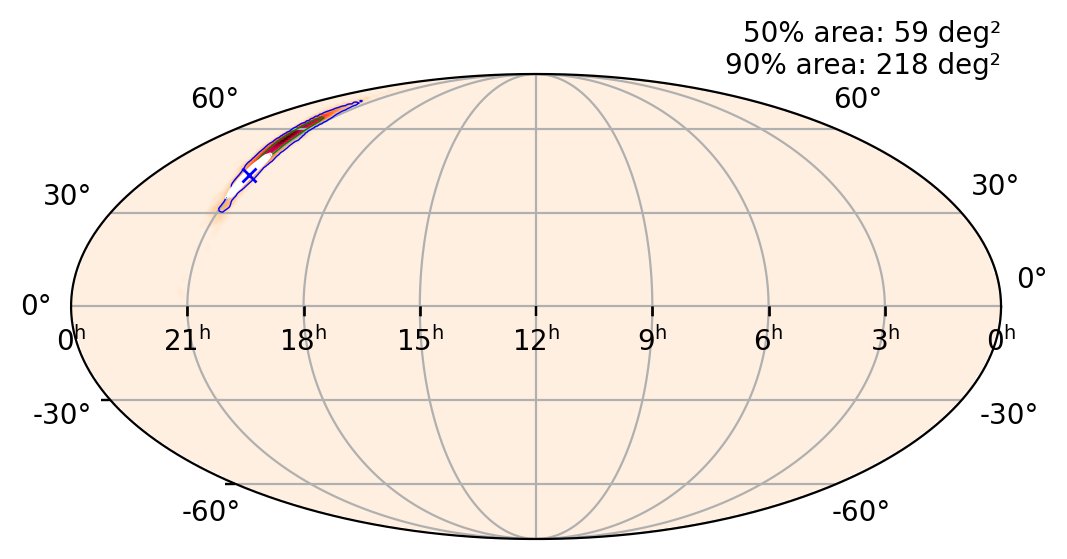}{0.45\textwidth}{(d)}
          \fig{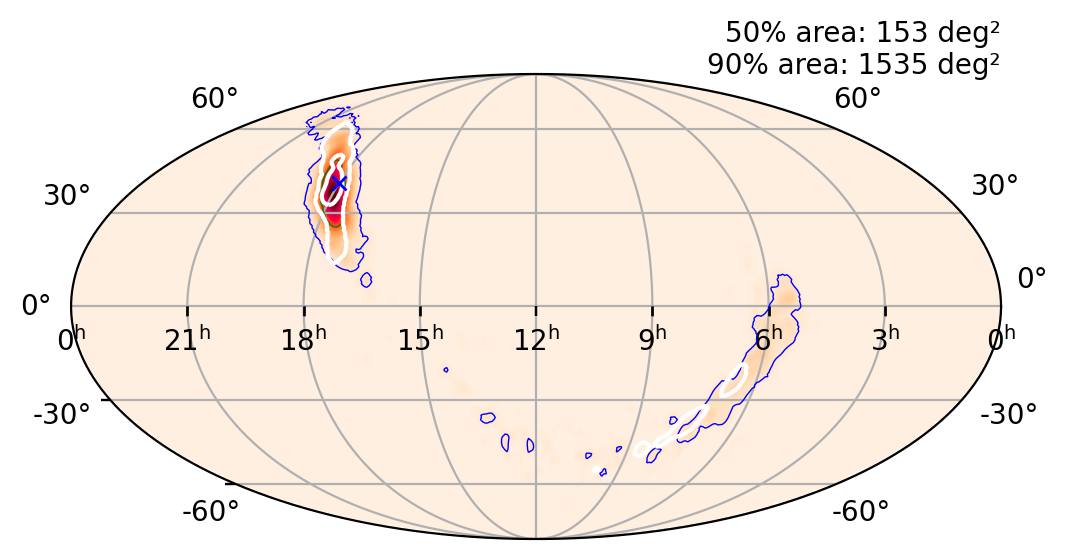}{0.45\textwidth}{(e)}}
\gridline{\fig{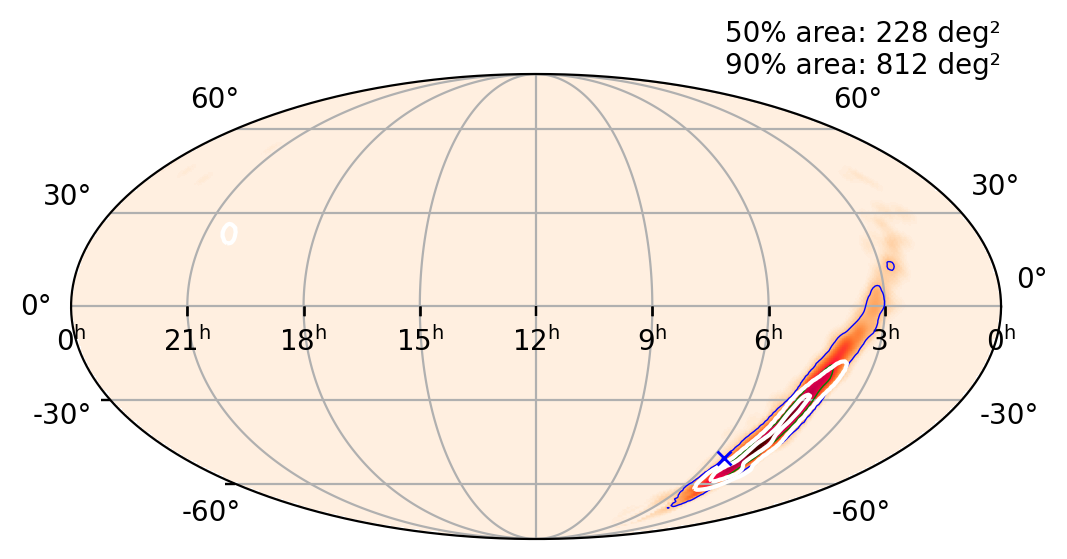}{0.45\textwidth}{(f)}}
\caption{Skymaps of three detector (upper panel) and two detector (bottom panel) BBH, BNS and NSBH (bottom panel) injections generated by \texttt{GW-SkyLocator}. The heatmaps show the probability density obtained from \texttt{GW-SkyLocator} with the 90\% and 50\% credible intervals show in blue and green respectively. The white contours shows the 90\% and 50\% credible regions obtained from \begin{small}BAYESTAR\end{small}. The blue marker show the true sky co-ordinates of the injections. The areas of the 90\% and 50\% credible regions from \texttt{GW-SkyLocator} are shown inset.}
\label{fig:3}
\end{figure}

\begin{figure}
\gridline{\fig{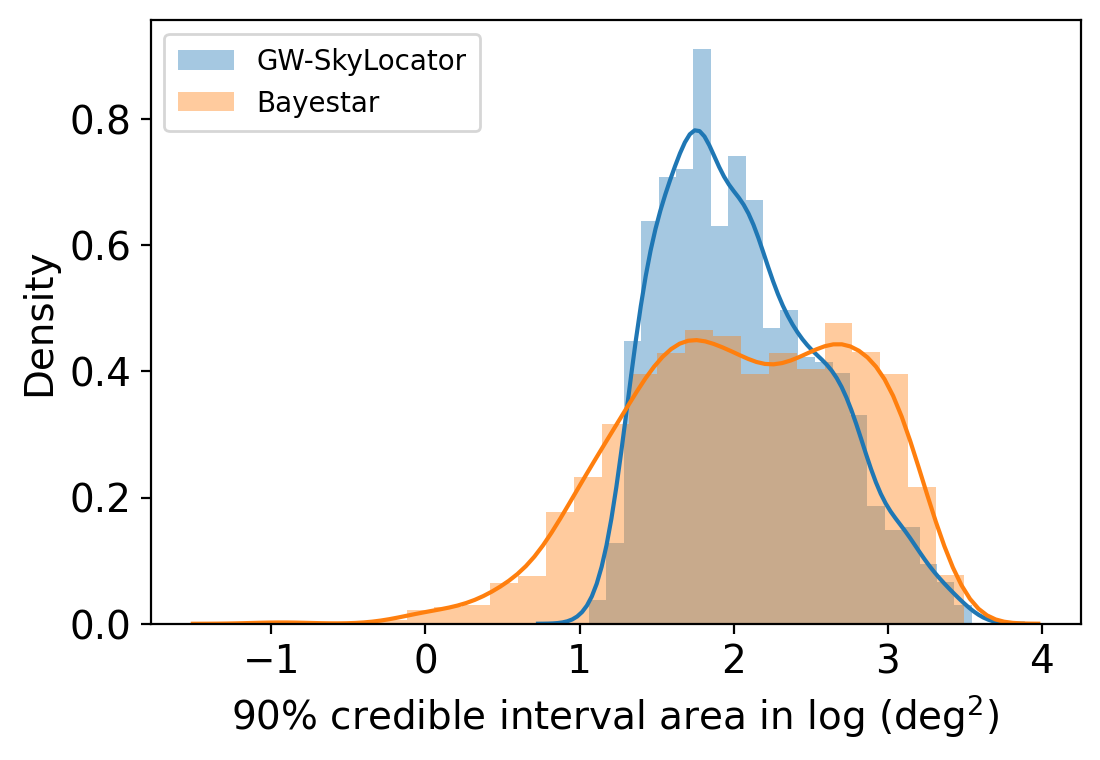}{0.32\textwidth}{(a)}
          \fig{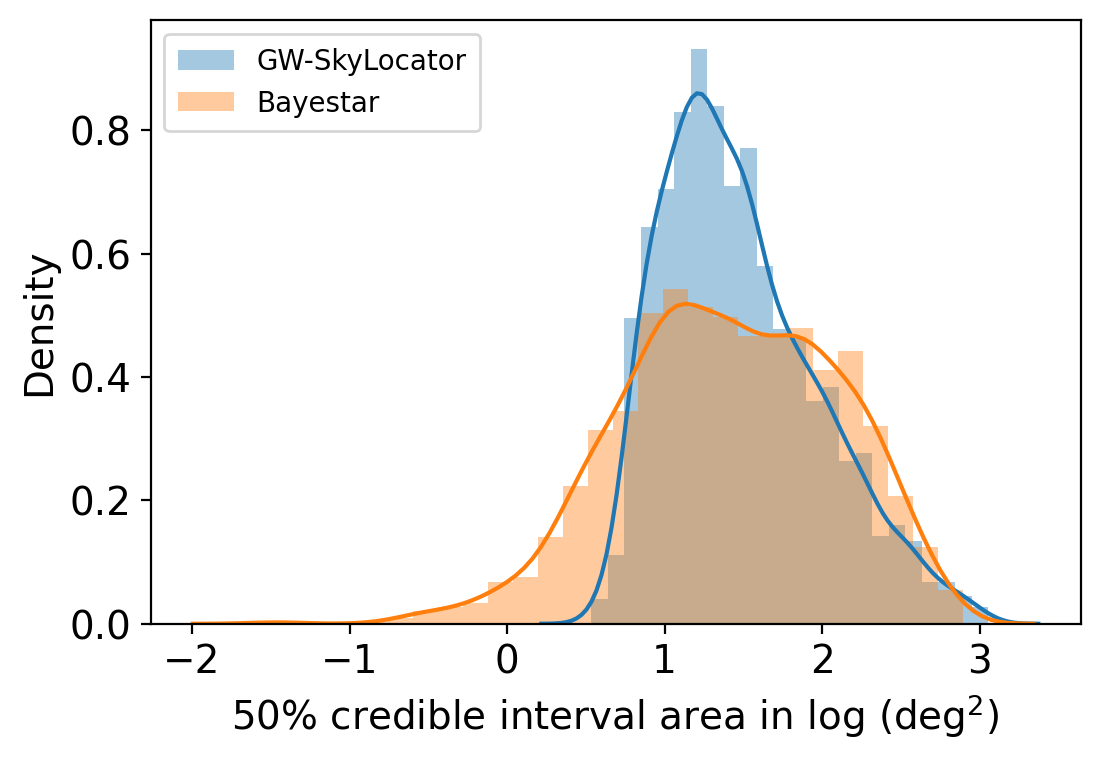}{0.32\textwidth}{(b)}
          \fig{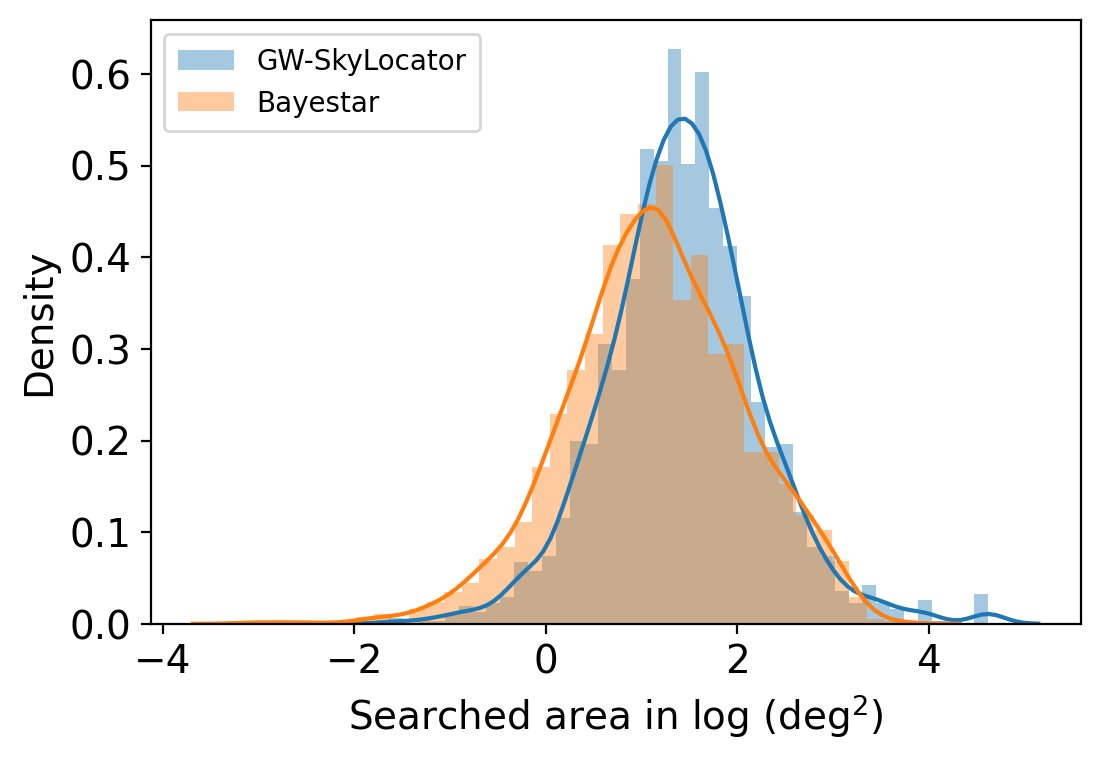}{0.32\textwidth}{(c)}}
\gridline{\fig{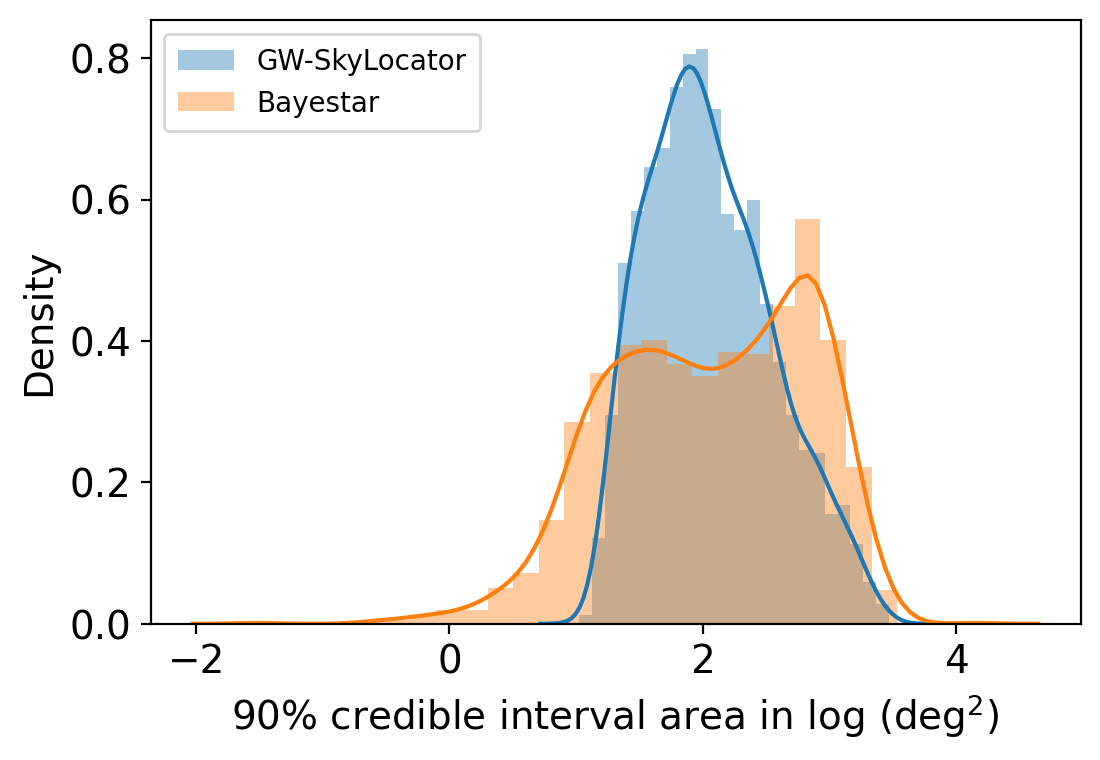}{0.32\textwidth}{(d)}
          \fig{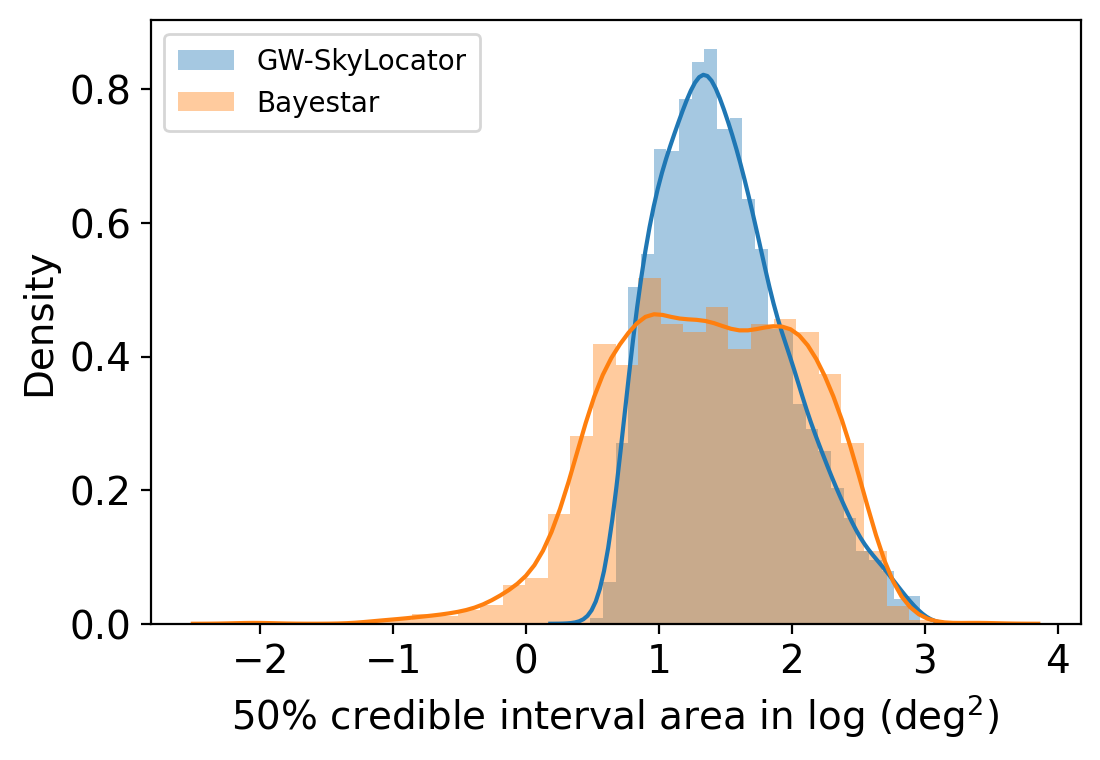}{0.32\textwidth}{(e)}
          \fig{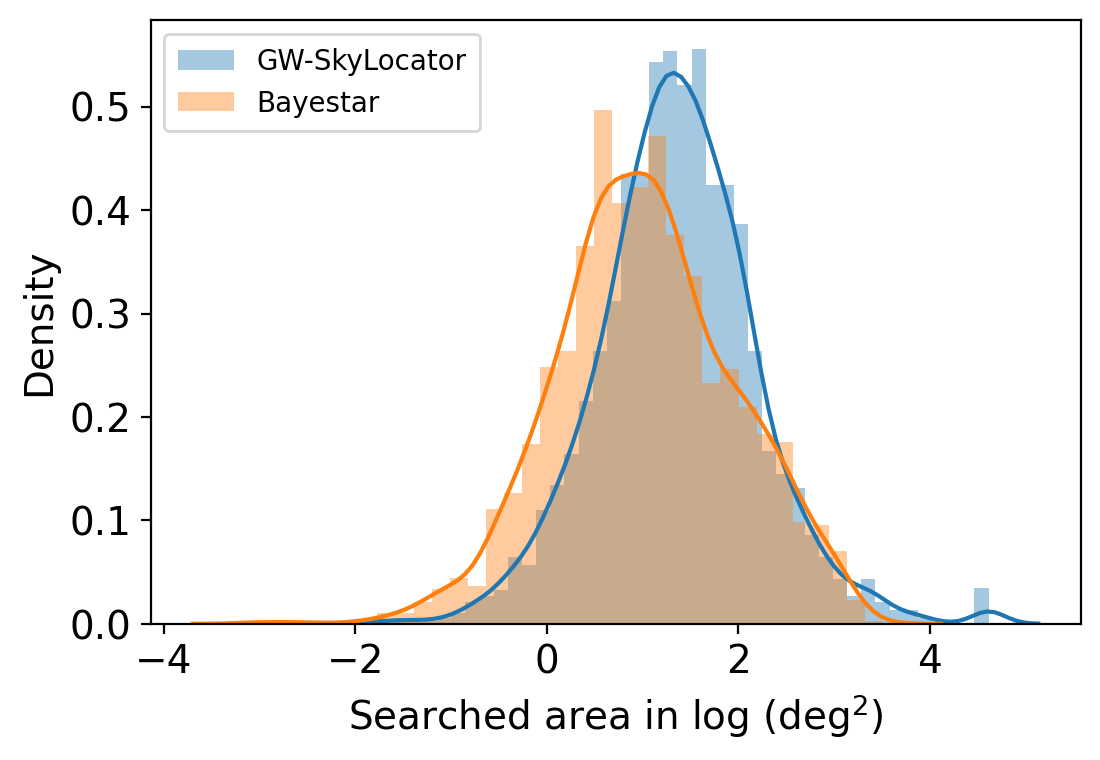}{0.32\textwidth}{(f)}}
\gridline{\fig{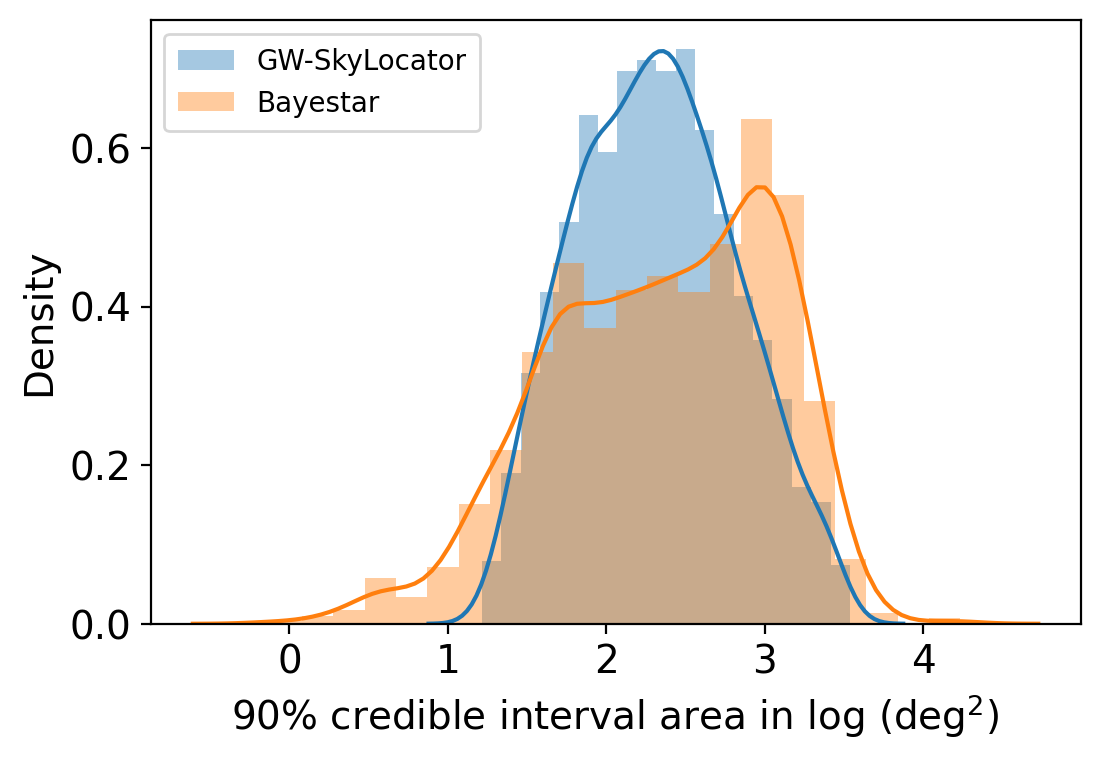}{0.32\textwidth}{(d)}
          \fig{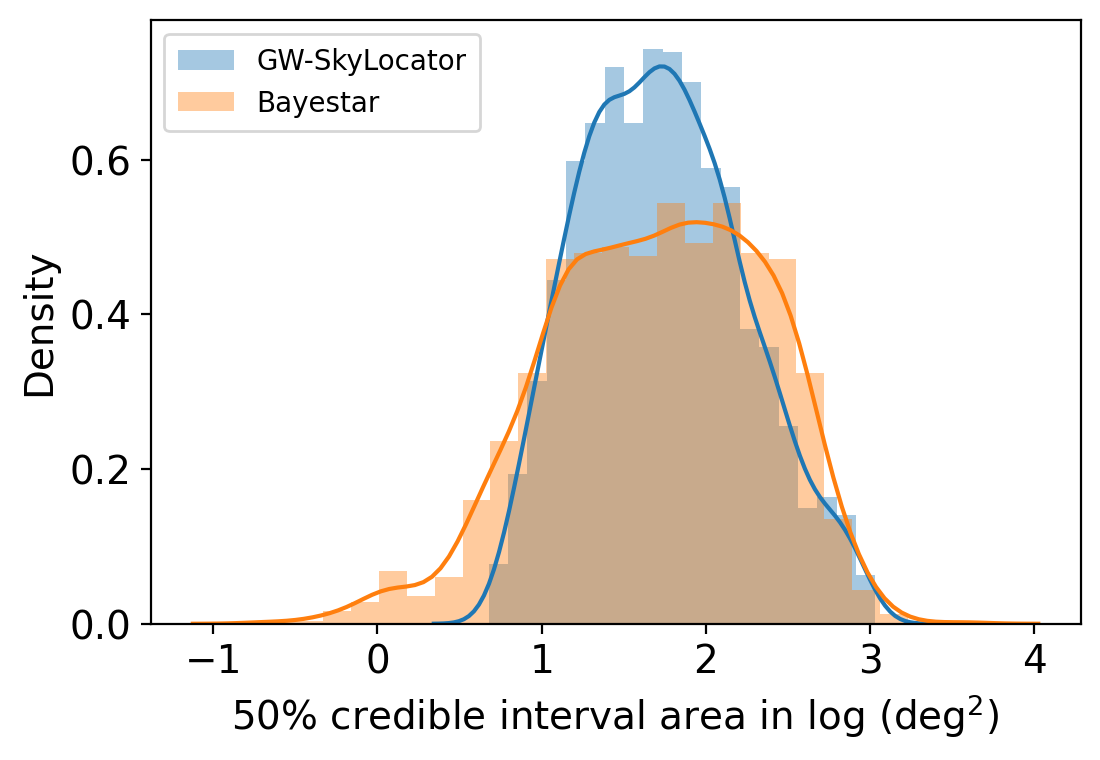}{0.32\textwidth}{(e)}
          \fig{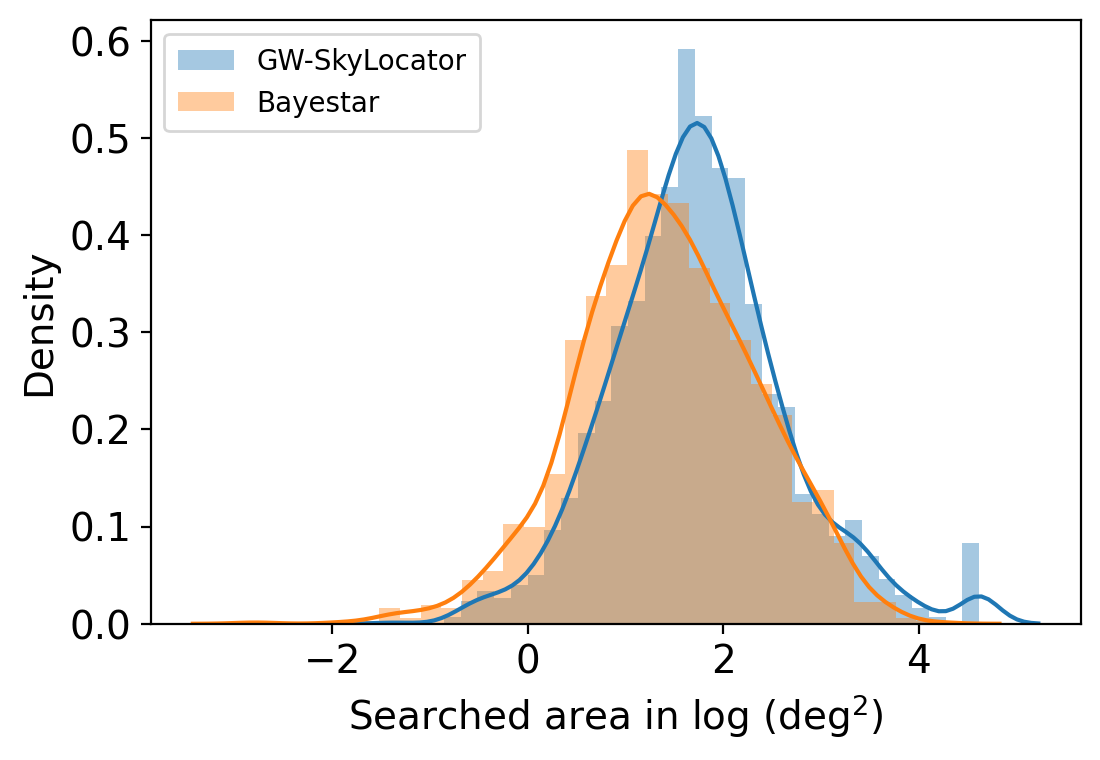}{0.32\textwidth}{(f)}}
\caption{\label{fig:Injection_run_3_det} Upper panel: (a) - (c) show densities of the areas of the 90\% credible intervals, 50\% credible intervals and searched areas respectively for our BBH injection set, obtained using \texttt{GW-SkyLocator} (blue) and \begin{small}BAYESTAR\end{small} (orange). Middle panel: (d) - (f) show densities of the 90\% credible interval areas, 50\% credible interval areas and searched areas for BNS samples. Bottom panel: (g), (h) and (i) show the same for NSBH injections.}
\label{fig:4}
\end{figure}

We discuss our model's performance on simulated GW events injected in stationary Gaussian noise with Advanced LIGO PSD. Figs.~\ref{fig:3} (a) -- (f) show \texttt{GW-SkyLocator}'s performance on BBH, BNS and NSBH injections respectively. The heatmaps in the figures show the probability densities of the source's sky location obtained by our model. The outer (blue) and inner (green) contours corresponding to the heatmaps in the figures represent the 90\% and 50\% credible interval regions respectively, and the blue marker shows the true sky co-ordinates of the injections. We observe that the true sky locations are within our model's 90\%  credible interval regions. In addition, we have also plotted the 90\% and 50\% credible intervals (white) obtained by \begin{small}BAYESTAR\end{small} for these injections. We notice while our 90\% credible intervals overlap with \begin{small}BAYESTAR\end{small}, the areas of \texttt{GW-SkyLocator}'s 90\% and 50\% intervals are usually larger than \begin{small}BAYESTAR\end{small}, especially for high S/N injections. This is also reflected in the histograms in Figs.~\ref{fig:4} (a)--(i) which show the densities of the areas of the 90\% credible intervals, 50\% credible intervals and searched areas from \texttt{GW-SkyLocator} and \begin{small}BAYESTAR\end{small} for 2000 BBH, BNS and NSBH injection samples detected in co-incidence by two or more detectors using matched filtering from a catalogue of simulated injections. \\

Figs.~ \ref{fig:4} (a), (d) and (g) show the distribution of the areas of the 90\% credible intervals of BBH, BNS and NSBH samples respectively, as obtained by \texttt{GW-SkyLocator} (blue) and \begin{small}BAYESTAR\end{small} (orange) respectively. We observe that for all three test cases, the median of \texttt{GW-SkyLocator}'s histograms lie close to the medians obtained by \begin{small}BAYESTAR\end{small}. The 90\% area histograms of \begin{small}BAYESTAR\end{small} for all three test cases show two distinct peaks, while the distribution obtained by our model shows a single peak centered in between the two peaks of \begin{small}BAYESTAR\end{small}. Therefore, on average, the areas of the 90\% credible regions obtained by the two methods are consistent. We however observed that for high S/N cases ($>$ 25), our model does not reproduce the same localization accuracy as \begin{small}BAYESTAR\end{small}, which is evident from the higher fraction of test samples with 90\% area $<$ 10 deg$^2$ in \begin{small}BAYESTAR\end{small} compared to \texttt{GW-SkyLocator} for all three cases. A possible reason for this is our choice of a fixed KDE bandwidth during inference to obtain the probability densities. While the LVK Collaboration uses the \texttt{ligo.skymap} \cite{Bayestar} software's implementation of a k-means clustering algorithm to find appropriate KDE bandwidths for each event separately, this method takes a few minutes to evaluate and therefore defeats our objective of rapid sky localization. We notice a similar trend in the distributions of the 50\% credible interval areas (Figs.~ \ref{fig:4} (b), (e) and (h) for BBH, BNS and NSBH respectively), in which the histograms from \texttt{GW-SkyLocator} and \begin{small}BAYESTAR\end{small} overlap for 50\% areas greater than $\sim$ 10 deg$^2$, but show a significant difference in the fraction of samples with areas less than 10 deg$^2$. The distributions of the searched areas (Figs.~\ref{fig:4} (c), (f) and (i) for BBH, BNS and NSBH respectively) for both \texttt{GW-SkyLocator} and \begin{small}BAYESTAR\end{small} however, are more Gaussian, with \begin{small}BAYESTAR\end{small}'s median searched area for all three cases being slightly smaller than that of \texttt{GW-SkyLocator}. \\

\begin{figure}
\begin{center}
     \includegraphics[scale=0.45]{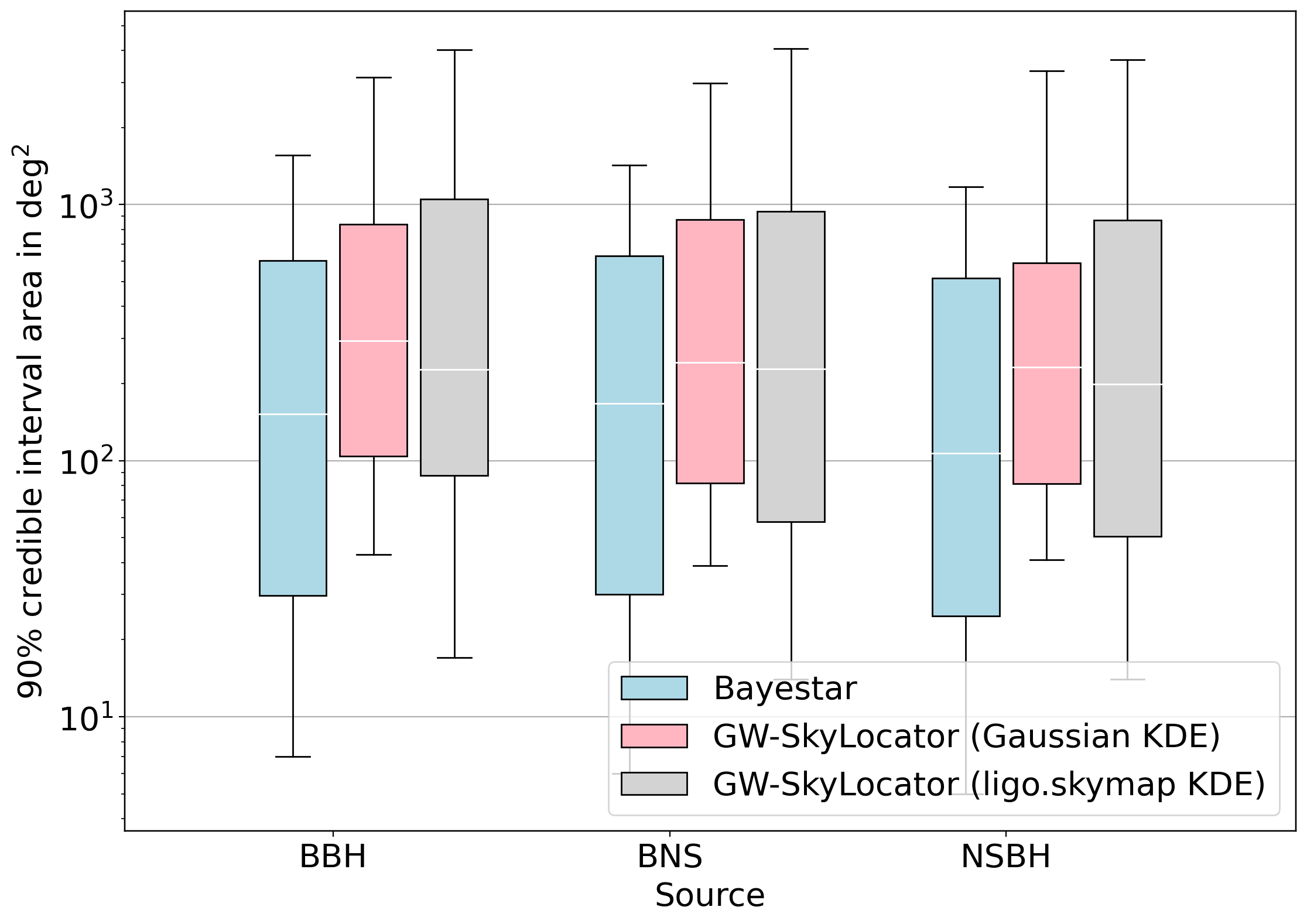}
\caption{Box and whisker plots showing areas of the 90\% credible intervals for a set of 200 BBH, BNS and NSBH injections each, obtained using \begin{small}BAYESTAR\end{small}, \texttt{GW-SkyLocator} with fixed bandwidth Gaussian KDE, and \texttt{GW-SkyLocator} with \texttt{ligo.skymap}'s KDE.}
\label{fig:6}
\end{center}
\end{figure}

We have investigated the effect of employing a fixed KDE bandwidth instead of \texttt{ligo.skymap}'s k-means clustering algorithm on the overall uncertainties of the sky location estimates, as shown in Figs.~\ref{fig:5} and \ref{fig:6}. Fig.~\ref{fig:5} (a) shows skymap for a simulated BBH event generated using \begin{small}BAYESTAR\end{small}. Fig. \ref{fig:5} (b) shows the same event's skymap generated using \texttt{GW-SkyLocator} with a fixed bandwidth Gaussian KDE and Fig.~\ref{fig:5} (c) shows result from \texttt{GW-SkyLocator} with \texttt{ligo.skymap}'s KDE implementation. Figs.~\ref{fig:5} (d) -- (f) and (g) -- (i) shows similar results for a simulated BNS and NSBH event respectively. We observe that implementing KDE with k-means clustering results in significant shrinkage of the sky location uncertainty, compared to what we obtain using a KDE with fixed bandwidth. We further run an injection test using 200 BBH, BNS and NSBH samples each and generate box and whisker plots for the 90\% credible interval areas obtained using \begin{small}BAYESTAR\end{small}, \texttt{GW-SkyLocator} with fixed bandwidth KDE and \texttt{GW-SkyLocator} with \texttt{ligo.skymap}'s KDE, as shown in Fig.~\ref{fig:6}. The improvement in the sky direction estimates from employing k-means clustering for bandwidth selection is apparent from the locations of the lower quartiles and the lower extreme tails of the box and whisker plots. While the smallest 90\% credible interval area obtained using a fixed KDE is around 40 deg$^{2}$, the minimum area with \texttt{ligo.skymap} KDE turns out to be less than 15 deg$^{2}$ for events in this injection set. Apart from the differences arising from our choice of KDE, Figs.~\ref{fig:4} -- \ref{fig:6} indicate that further optimization of our networks is needed to match the overall accuracy of \begin{small}BAYESTAR\end{small}. 

\begin{figure}
\gridline{\fig{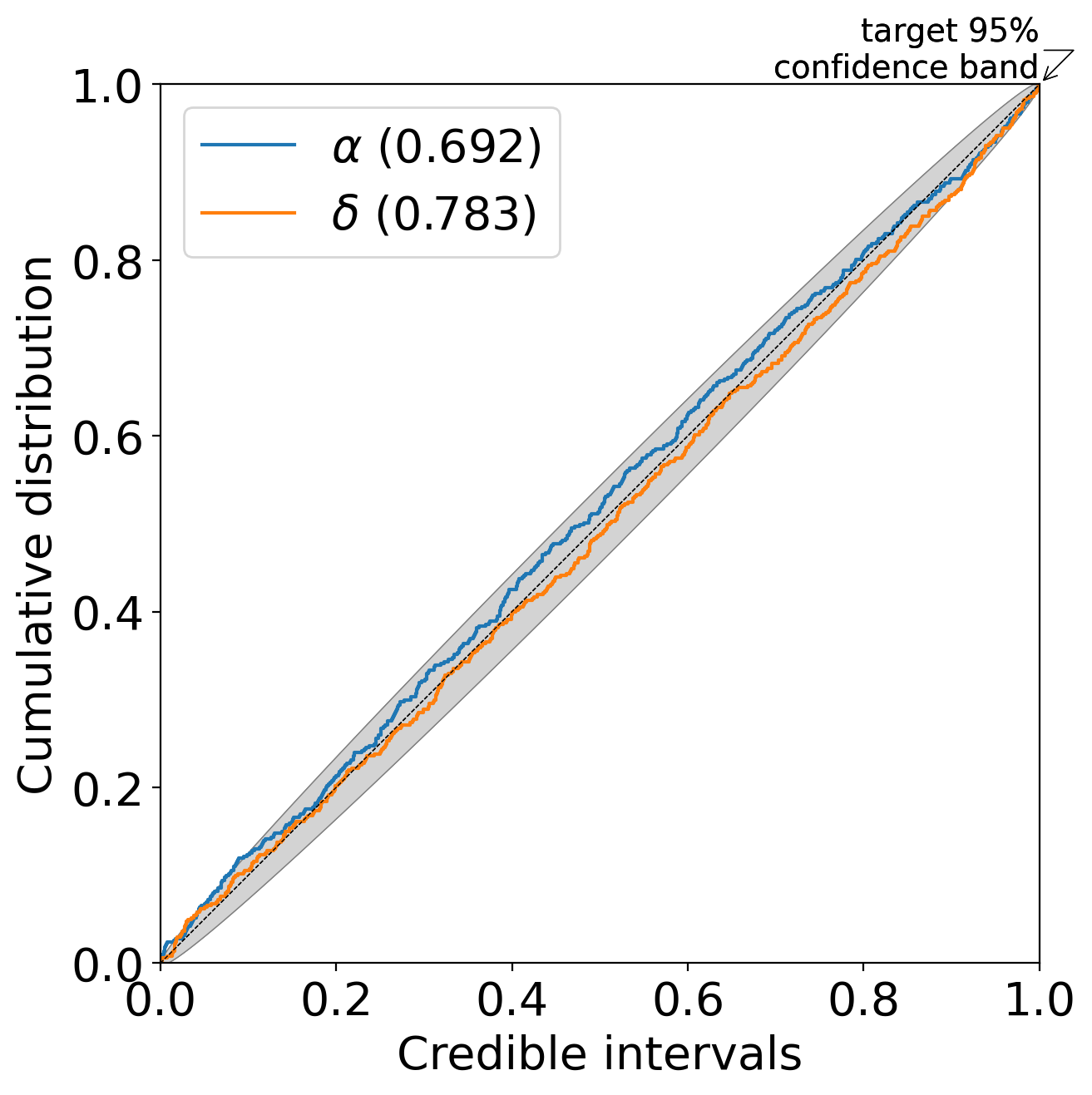}{0.32\textwidth}{(a)}
          \fig{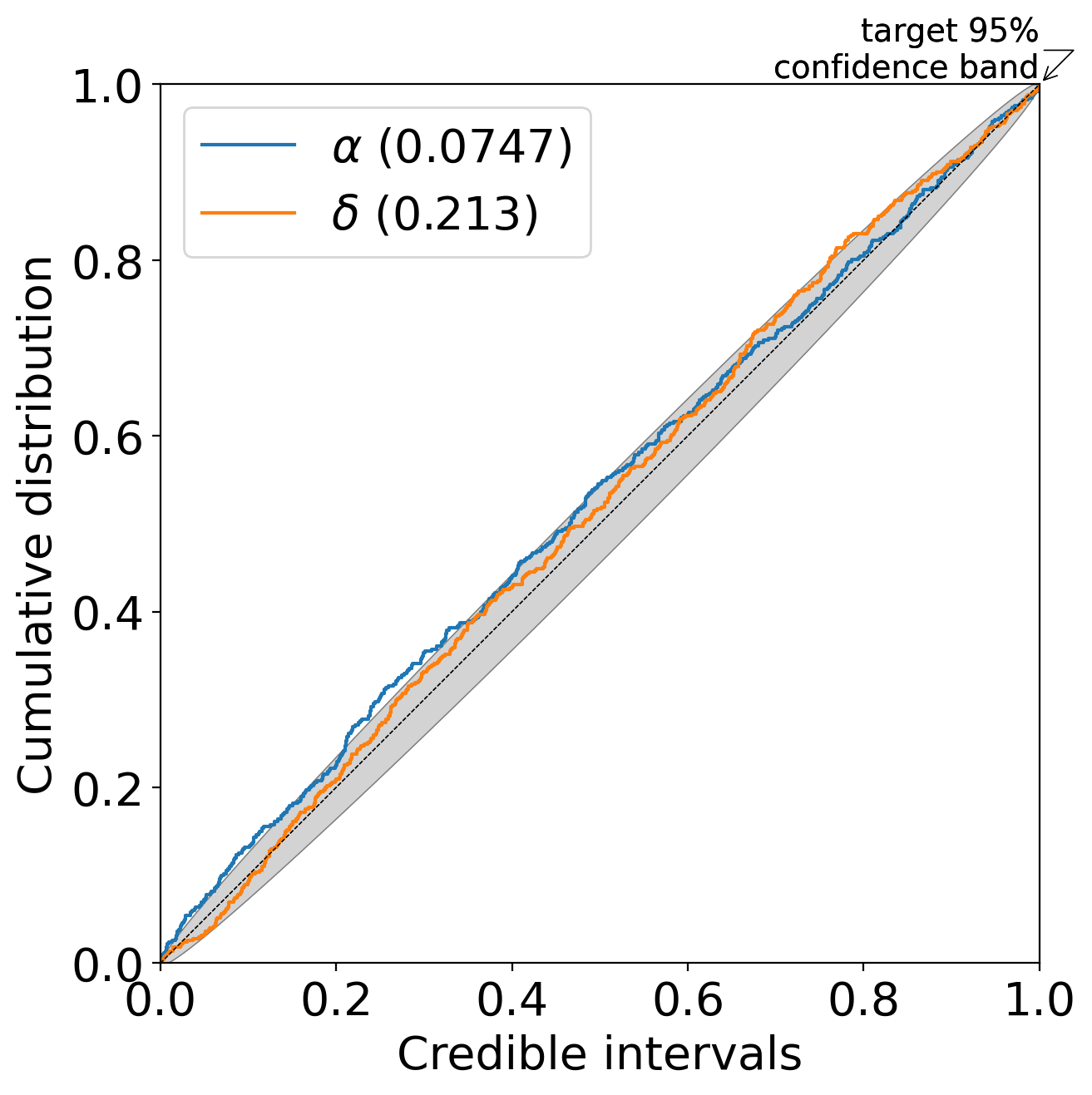}{0.32\textwidth}{(b)}
          \fig{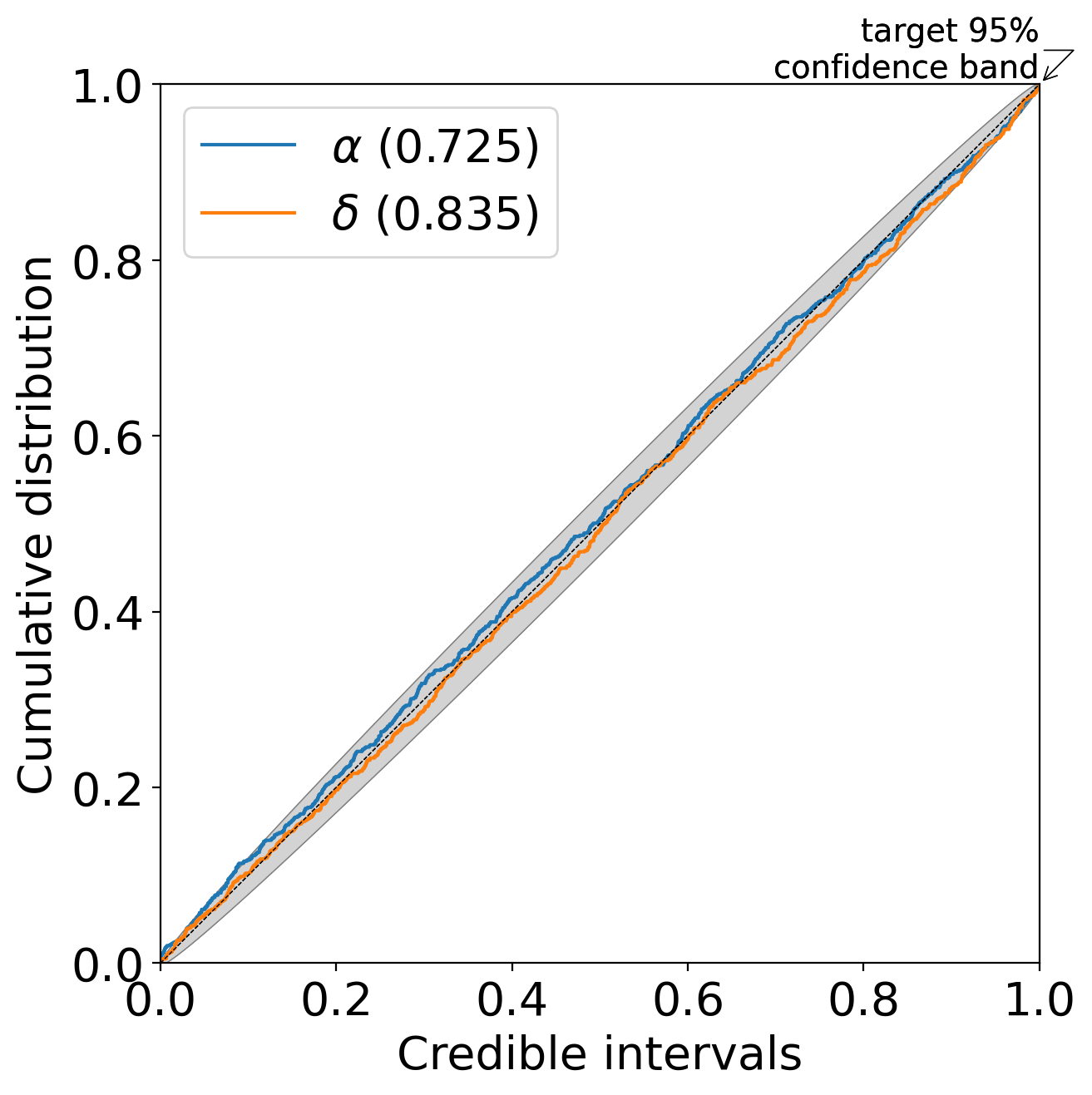}{0.32\textwidth}{(c)}}
\caption{Left to right: P–P plots for BBH, BNS and NSBH test samples respectively. The dotted diagonal line shows expected cumulative distribution of events within different credible intervals. The grey bands in the P-P plots show the 95\% errors margins arising due to
statistical fluctuations. The p-values for the KS tests of $\alpha$ and $\delta$ for each injection is shown in the legend.}
\label{fig:7}
\end{figure}

Figs.~\ref{fig:7} (a), (b) and (c) show Probability-Probability (P-P) plots for \texttt{GW-SkyLocator}'s predictions on the BBH, BNS and NSBH injections respectively. The P-P plots show the cumulative histograms of the true sky directions of events lying in different credible intervals. It serves as a test of consistency of any Bayesian parameter estimation method with the frequentist interpretation (i.e., on average the true sky directions should lie within the 90\% interval for 90\% of the test cases). To generate the P-P plot, for each skymap, we rank the sky pixels by descending probability density and sum up their probabilities until we reach the pixel that contains the true sky location for the injection. We then plot the empirical cumulative distribution of these injections. For accurate posteriors, the cumulative distribution should be a diagonal line, i.e., it should follow the dotted lines in Figs.~\ref{fig:7} (a), (b) and (c). From the figures we observe that our empirical distribution lies within the 95\% confidence interval of the dotted lines. This is also evident from the p-values of the Kolmogorov-Smirnov (KS) tests we performed for $\alpha $ and $\delta $ (see inset in Figs.~\ref{fig:7} (a), (b) and (c)). The results of the KS tests (p values  $>$ 0.05) suggest that at a 5\% level of significance, we cannot reject the null hypothesis that the cumulative distributions of the percentile scores of the true sky directions within their marginalized credible intervals is uniform, i.e., they follow the diagonal line. These tests suggest that our posteriors are indeed accurate.

\subsection{Localization of real GW events}

We tested our model's performance on real GW events detected during the second and third observation runs of LIGO and Virgo and compared our model's performance with published results from \begin{small}BAYESTAR\end{small} and LALI\begin{small}NFERENCE\end{small}. In order to test on real events, we retrained our model on injections in real LIGO-Virgo data from the second and third observation runs. This was done because the PSD of Advanced LIGO, which was used for the previous analyses, is significantly different from that of the real events. Therefore, the model had to be retrained with S/N time series data obtained from similar PSDs in order to obtain accurate sky location distributions. For O3 data, we used real noise from GPS times between 1242181632 and 1242705920, and for O2 data we used noise between GPS times 1187008512 and 1187733618. The noise and the GW strain data for the real events was downloaded from the public data release available at the Gravitational Wave Open Science Center (GWOSC). We analyzed the strain data of the real events and obtained their respective S/N time series by performing matched filtering using templates with source parameters whose values were drawn from the medians of the published O2 and O3 source parameter estimation results (\cite{GWTC3}). It is to be noted that while our model is trained on the optimal templates S/N time series, which is obtained by matched filtering strains with the exact templates, for real events the S/N time series are not optimal since the true values of the source parameters are unknown. \\

\begin{figure}
\gridline{\fig{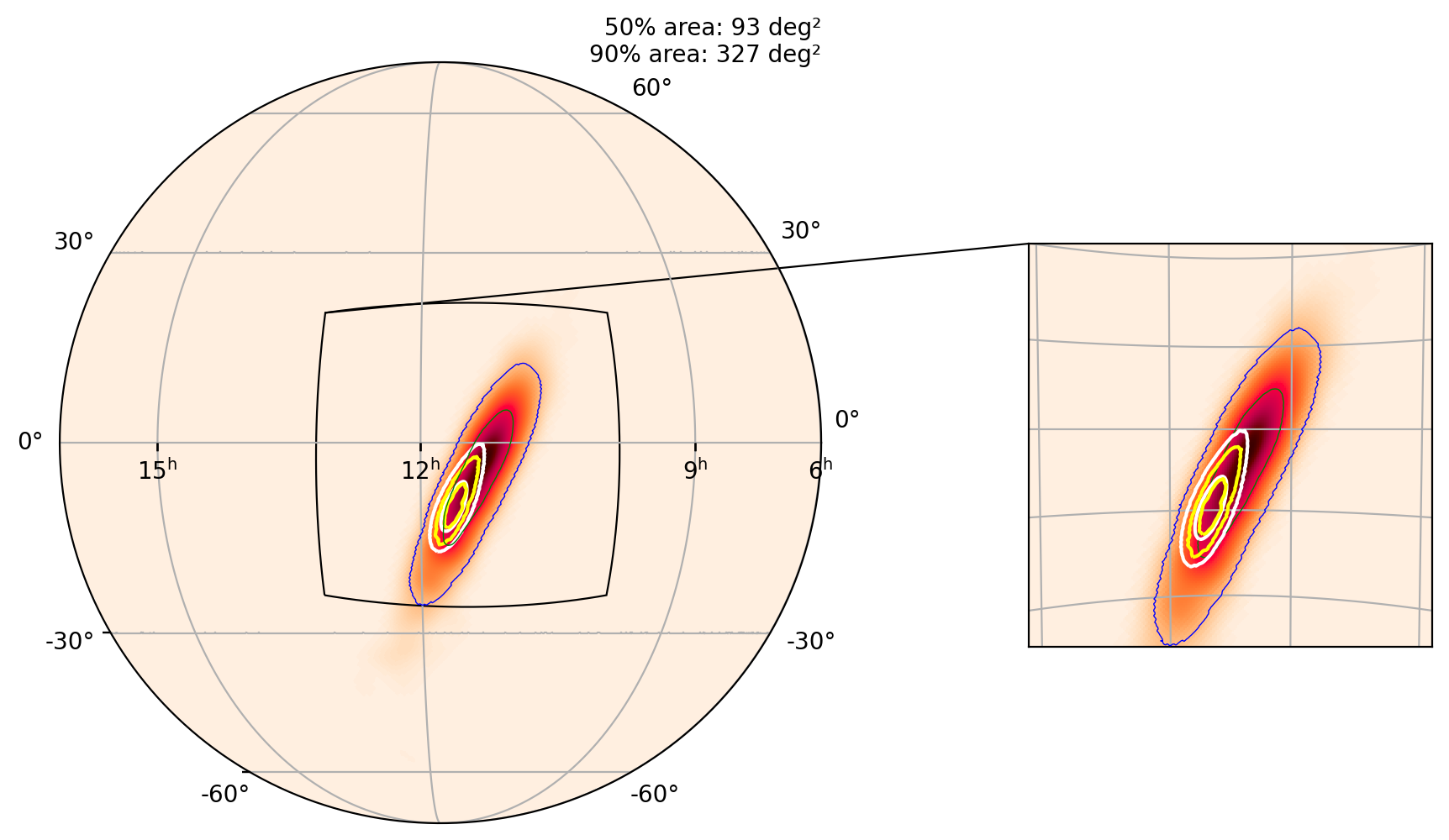}{0.45\textwidth}{(a)}
          \fig{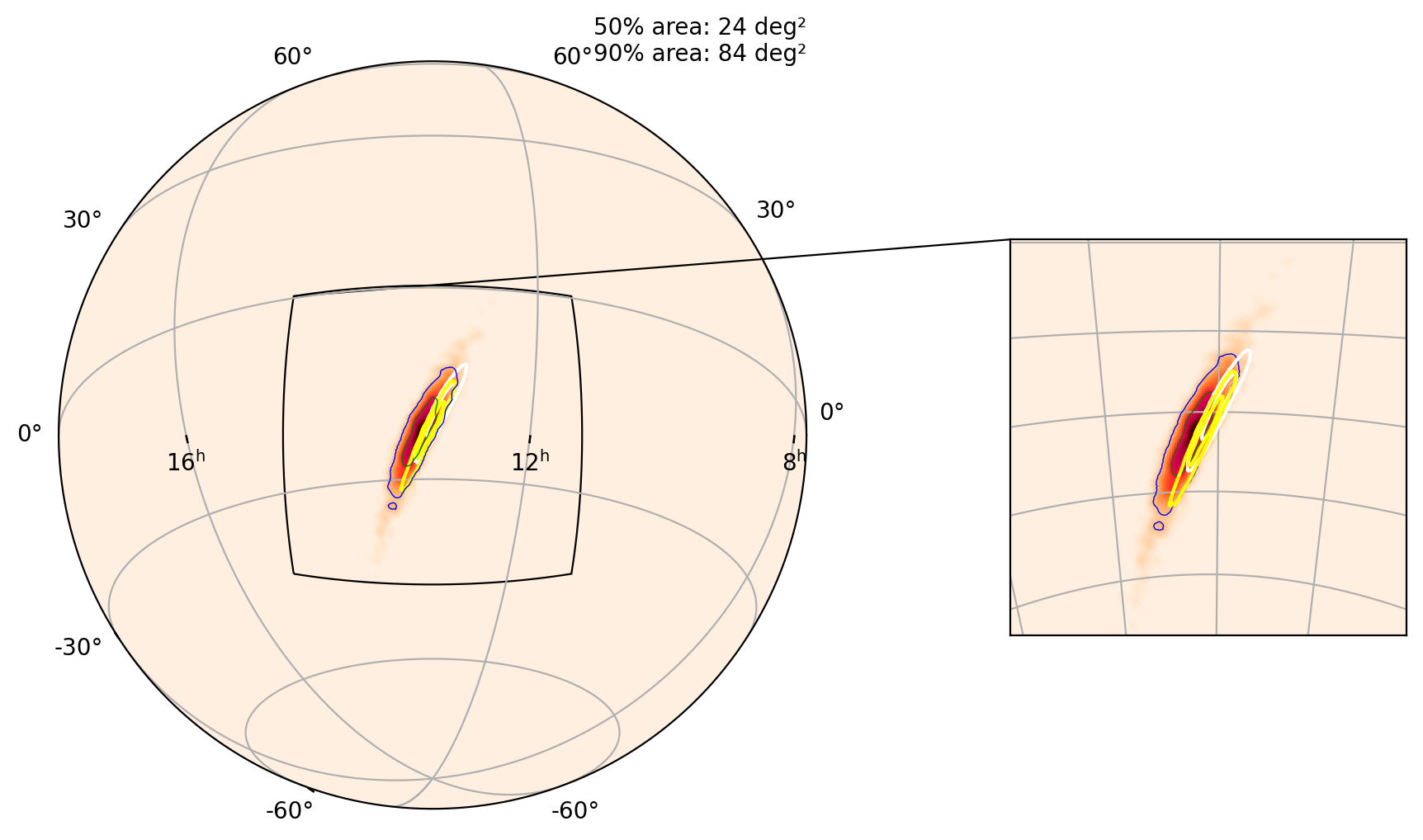}{0.45\textwidth}{(b)}}
\gridline{\fig{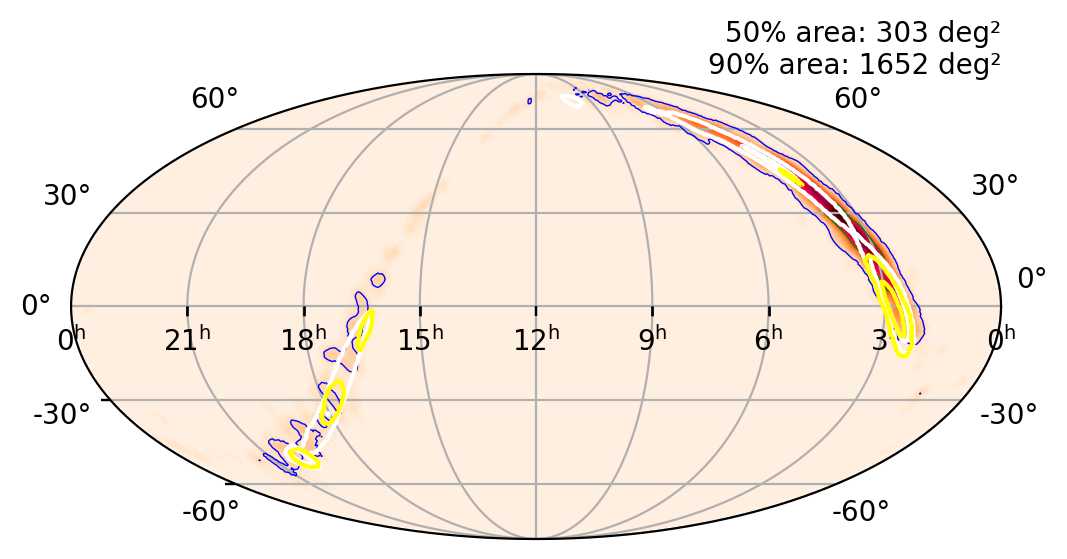}{0.45\textwidth}{(c)}}
\caption{Skymaps of real GW events detected during O2 and O3 generated by \texttt{GW-SkyLocator}. The upper panel shows skymaps for the BBH and BNS events GW200224\_22234 and GW170817 respectively. The bottom panel shows skymap for the NSBH event GW200115\_042309. The heatmaps show the probability density obtained from \texttt{GW-SkyLocator} with the 90\% and 50\% credible intervals show in blue and green respectively. The yellow contour shows published sky localization results obtained from full parameter estimation software and the white contour shows the sky location estimates obtained from \begin{small}BAYESTAR\end{small}. The areas of the 90\% and 50\% credible regions from \texttt{GW-SkyLocator} are shown are shown inset.}
\label{fig:8}
\end{figure}

Fig.~\ref{fig:8} (a), (b) and (c) show skymaps for the BBH event  GW200224\_2223 detected during the second half of the third observation run (O3b), the first BNS event detected by LIGO and Virgo, GW170817, and the NSBH event GW200115\_042309 detected during the third observation run. These events were selected in particular because their masses and distances fall within our chosen prior ranges for generating the training sets. We have also plotted the 90\% and 50\% credible intervals from \begin{small}BAYESTAR\end{small} (white) and LALI\begin{small}NFERENCE\end{small} (yellow) for comparison. We notice that our predicted contours overlap well with these methods but are larger in area. A possible explanation for this, which is related to our choice of fixed KDE bandwidth has already been provided. Another explanation for this mismatch could be the difference in the PSDs between our training and test sets. While, we generated training data from PSDs drawn from O2 and O3, real LIGO-Virgo data is non-stationary, and therefore the PSDs corresponding to each individual event is different. In order to make our model robust against different PSDs, it is essential to condition the Normalizing Flow on the PSD computed from strain data close to the events being tested on. This method was applied in a recent study by Dax et al. (2022) (\cite{Green}) to obtain unprecedented parameter estimation accuracy of real BBH events from O1 and O2. We plan to incorporate this feature in our network in the future. \\

We also notice that the shape of the localization contours for GW200224\_22234 and GW170817 are different from GW200115\_042309. This is because the shape of the sky localization contours is primarily determined by the signal arrival time delays, and is partially modulated by the phase lag and amplitude difference between signals at individual detectors. For each pair of detectors, one can plot a ring of possible sky directions, for a single time of arrival difference. For a three detector network, the rings intersect at two patches in the sky which are mirror images with respect to the plane defined by the detector network. This degeneracy can be broken by considering the amplitude and phase consistency of the signals between detectors, which allows us to neglect one of the patches. For events like GW200115\_042309 in which the Virgo S/N is very weak, we obtain elongated ellipses and rings in the sky, which are only partially broken by including the amplitude and phase information, as we observe in Fig~ \ref{fig:8} (c). For GW170817, the observation of weak Virgo S/N however helped constrain the sky localization contour better by eliminating the region of the sky that Virgo has high sensitivity in, producing a highly constrained sky location posterior as a result. \\

\begin{figure}
\gridline{\fig{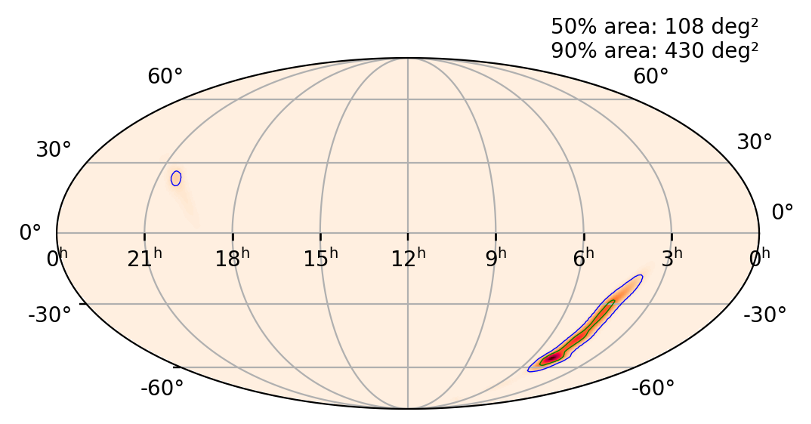}{0.32\textwidth}{(a)}
          \fig{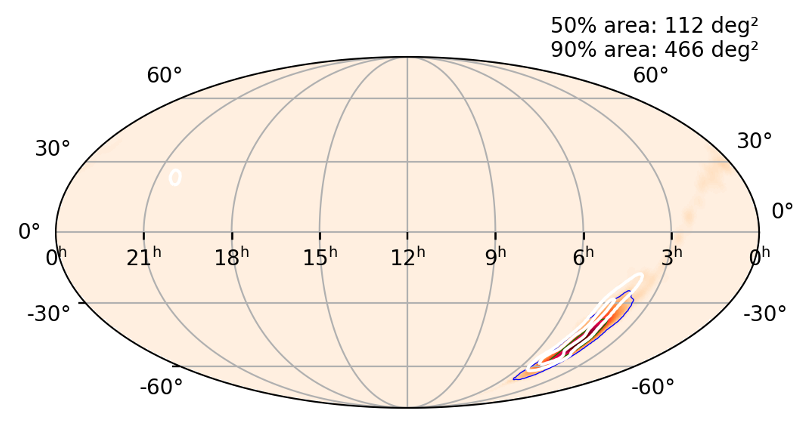}{0.32\textwidth}{(b)}
          \fig{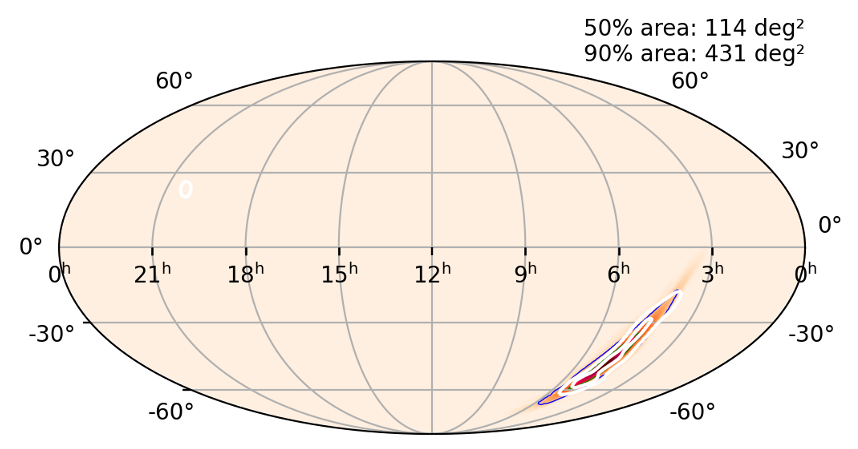}{0.32\textwidth}{(c)}}
\gridline{\fig{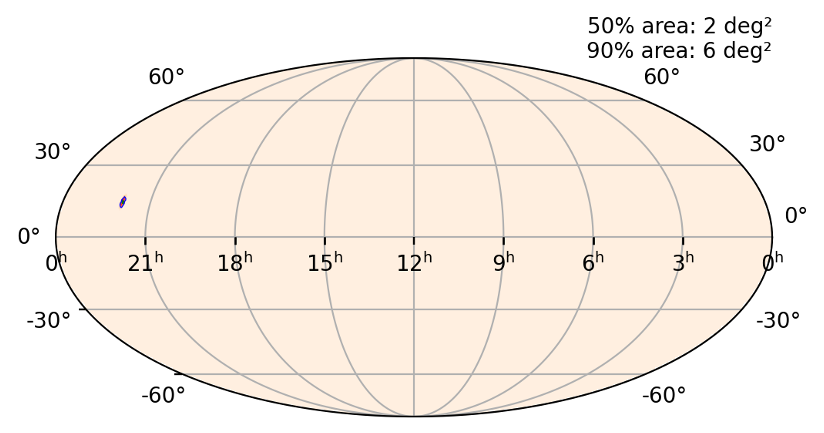}{0.32\textwidth}{(d)}
          \fig{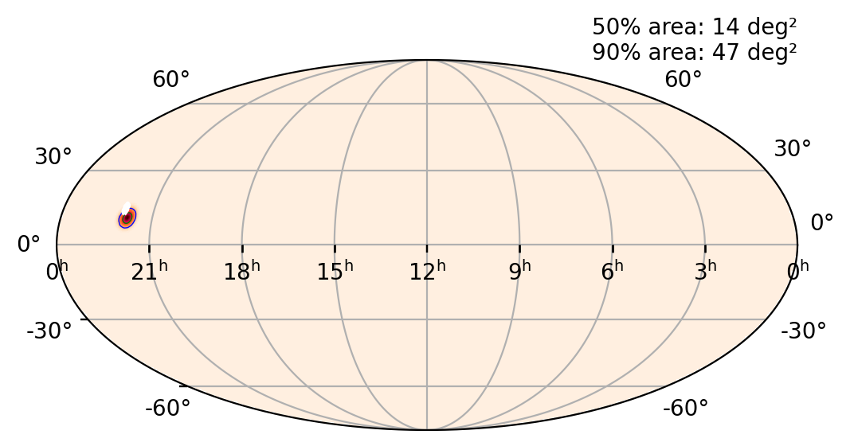}{0.32\textwidth}{(e)}
          \fig{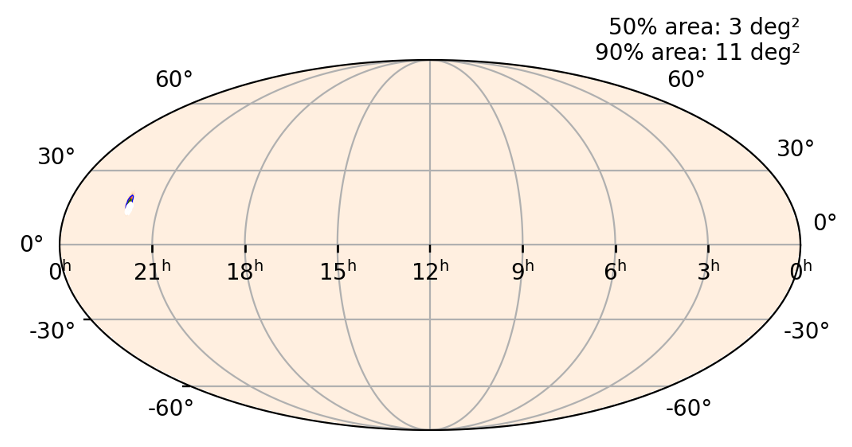}{0.32\textwidth}{(f)}}
\gridline{\fig{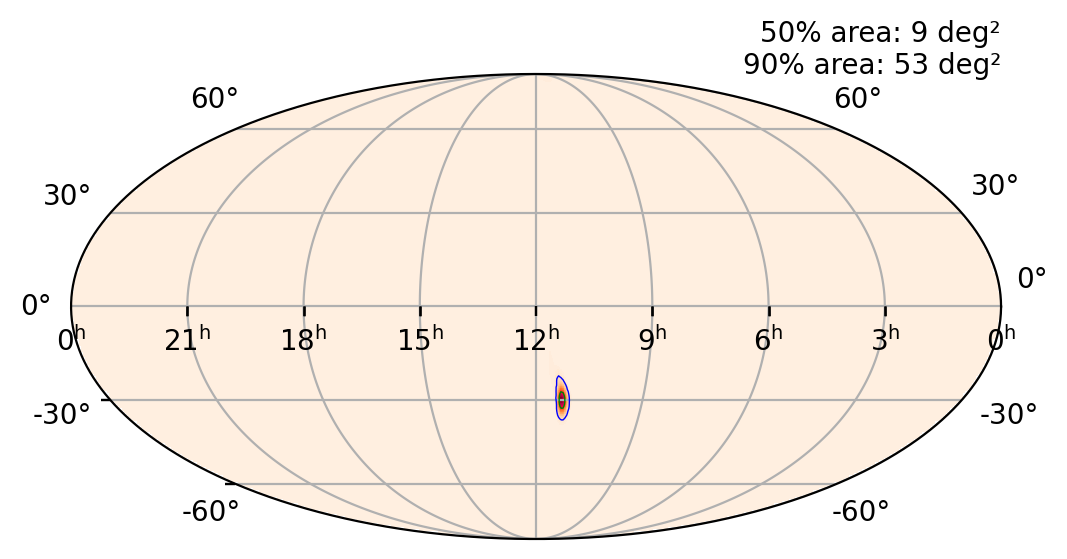}{0.32\textwidth}{(d)}
          \fig{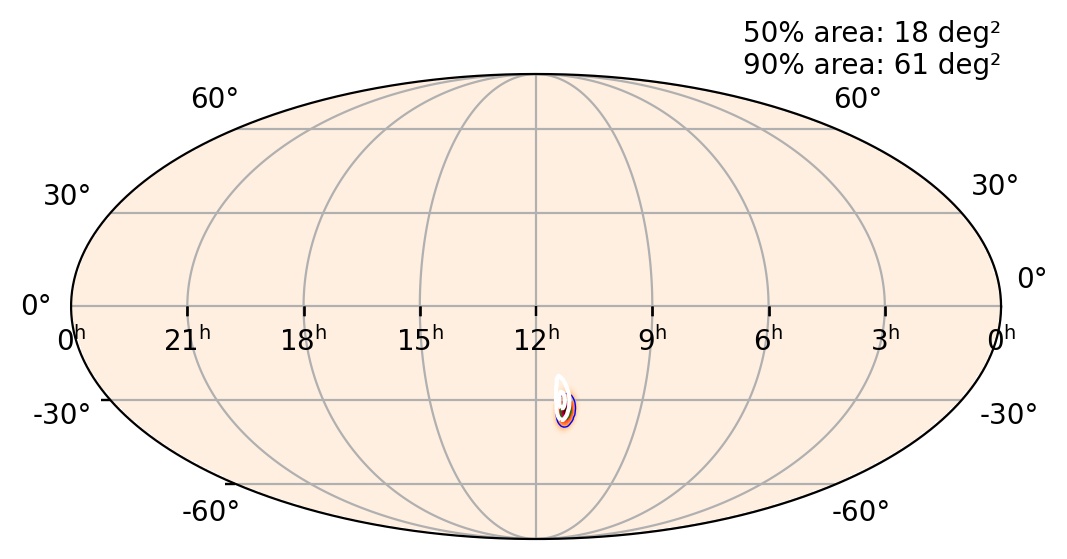}{0.32\textwidth}{(e)}
          \fig{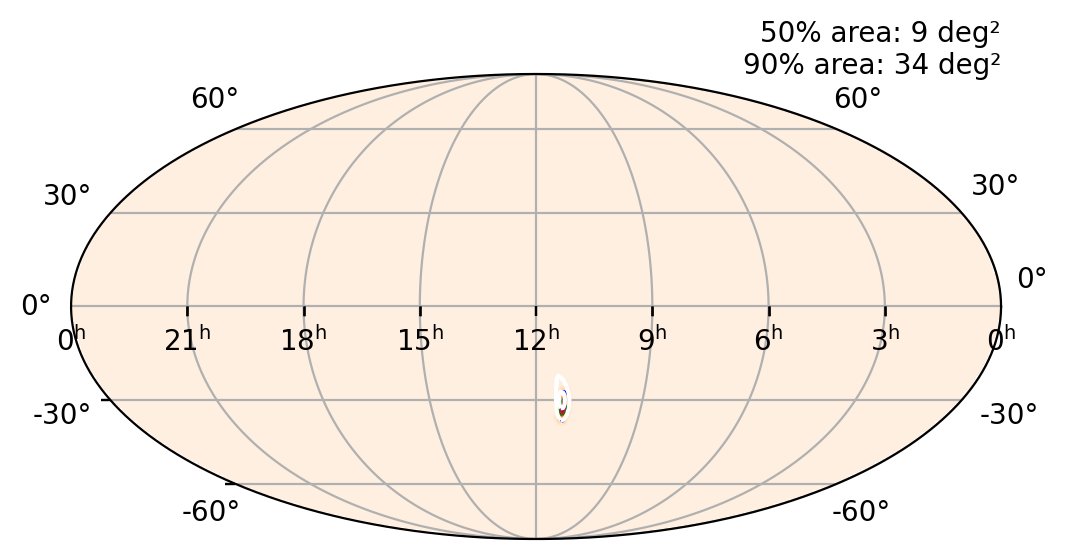}{0.32\textwidth}{(f)}}
\caption{\label{fig:5} Upper panel: (a) -- (c) show skymaps of a simulated BBH event obtained using \begin{small}BAYESTAR\end{small}, \texttt{GW-SkyLocator} with fixed bandwidth Gaussian KDE and \texttt{GW-SkyLocator} with \texttt{ligo.skymap}'s KDE respectively. Middle panel: (d) -- (f)  and bottom panels: (g) -- (i) show similar skymaps for simulated BNS and NSBH events obtained using the same methods as (a) -- (c).}
\label{fig:5}
\end{figure}

\section{Discussion}
In summary, we have reported the first deep learning based approach for sky localization of all CBC sources that enables orders of magnitude faster inference than other methods. Using S/N time series data instead of the actual GW strain data for training and testing, we were able to achieve localization of BNS and NSBH sources for the first time using deep learning. The use of S/N time series data also significantly reduces computational overheads and memory, which acts as a bottleneck in many deep learning-based approaches. We plan to adapt this method for extensively training and testing on real noise injections in the future. \textbf{Since deep learning methods do not require a likelihood to be explicitly defined, and can learn the distribution of data directly from training examples, they can possibly be more robust against glitches, non-stationary and non-Gaussian noise than other methods like \begin{small}BAYESTAR\end{small}, BILBY and LAL\begin{small}INFERENCE\end{small}, which assume a Gaussian likelihood. The present work is a feasibility study for deep learning as an alternative tool for rapid GW sky localization and it is hoped that certain modifications to the current model architecture will improve the consistency of our model's results to standard approaches, while having the same inference speed as \begin{small}BAYESTAR\end{small}. One of these modifications is to condition the network not only on the S/N time series and intrinsic parameters, but also on the individual detector PSDs of the noise around the event in consideration. This modification will make the inputs to our model exactly similar to \begin{small}BAYESTAR\end{small}'s (i.e., S/N time series, intrinsic parameters of best-matched templates, and detector PSDs), and it has also proven to significantly improve the performance of deep learning-based parameter estimation approaches in previous studies (\cite{Green}). We will investigate this feature in our future work.} In the future, we also plan to extend this method for rapid inference of the luminosity distance of GW sources, which is essential for prompt identification of the host galaxies for electromagnetic follow-up observations of compact object mergers.

\bibliography{bibliography}{}
\bibliographystyle{aasjournal}



\section{Acknowledgments}
This research was supported in part by the Australian
Research Council Centre of Excellence for Gravitational
Wave Discovery (OzGrav, through Project No. CE170100004). This research was undertaken with the support of computational resources from the Pople high-performance computing cluster of the Faculty of Science at the University of Western Australia.
This work used the computer resources of the OzStar computer cluster at Swinburne University of Technology. The OzSTAR program receives funding in part from the Astronomy National Collaborative Research Infrastructure Strategy
(NCRIS) allocation provided by the Australian Government. This research used data obtained from the Gravitational Wave Open Science Center (https://www.gw-openscience.org), a service of LIGO Laboratory, the LIGO Scientific Collaboration and the Virgo Collaboration. LIGO is funded by the U.S. National Science Foundation. Virgo is funded by the French Centre National de Recherche Scientifique (CNRS), the Italian Istituto Nazionale della Fisica Nucleare (INFN) and the Dutch Nikhef, with contributions by Polish and Hungarian institutes. This material is based upon work supported by NSF's LIGO Laboratory which is a major facility fully funded by the National Science Foundation. The authors would like to thank Prof. Amitava Datta from The University of Western Australia for his help in this work.

\end{document}